%% file: main.tex
\documentclass[sigconf]{acmart}

\usepackage{xspace}
\usepackage{subfigure}

\AtBeginDocument{%
  }

\setcopyright{acmcopyright}
\copyrightyear{2023}
\acmYear{2023}
\setcopyright{acmlicensed}\acmConference[FAccT '23]{2023 ACM Conference on Fairness, Accountability, and Transparency}{June 12--15, 2023}{Chicago, IL, USA}
\acmBooktitle{2023 ACM Conference on Fairness, Accountability, and Transparency (FAccT '23), June 12--15, 2023, Chicago, IL, USA}
\acmPrice{15.00}
\acmDOI{10.1145/3593013.3594098}
\acmISBN{979-8-4007-0192-4/23/06}

\newcommand\CSLINE{
  \global\let\savedtextbullet\textbullet
  \gdef\textbullet{%
    \par\noindent\savedtextbullet\global\let\textbullet\savedtextbullet
  }%
}
\begin{document}

\title{Auditing Cross-Cultural Consistency of Human-Annotated Labels for Recommendation Systems}

\author{Rock Yuren Pang}
\authornote{Both authors contributed equally.}
\authornote{Affiliated with Microsoft Research while research was conducted.}
\email{ypang2@cs.washington.edu}
\affiliation{%
  \institution{University of Washington}
  \country{Seattle, WA, USA}
}

\author{Jack Cenatempo}
\authornotemark[1]
\email{v-jcenatempo@microsoft.com }
\affiliation{%
  \institution{Microsoft Research}
  \country{Cambridge, MA, USA}
}

\author{Franklyn Graham}
\email{fgraha@gmail.com}
\affiliation{%
  \institution{Microsoft Corp.---Xbox Division}
  \country{Redmond, WA, USA}
}

\author{Bridgette Kuehn}
\email{bridgette@xbox.com}
\affiliation{%
  \institution{Microsoft Corp.---Xbox Division}
  \country{Redmond, WA, USA}
}

\author{Maddy Whisenant}
\email{mawhisen@xbox.com}
\affiliation{%
  \institution{Microsoft Corp.---Xbox Division}
  \country{Redmond, WA, USA}
}

\author{Portia Botchway}
\email{Portia.Botchway@microsoft.com}
\affiliation{%
  \institution{Microsoft Corp.---Xbox Division}
  \country{Redmond, WA, USA}
}

\author{Katie Stone Perez}
\email{kstone@microsoft.com}
\affiliation{%
  \institution{Microsoft Corp.---Xbox Division}
  \country{Redmond, WA, USA}
}

\author{Allison Koenecke}
\authornotemark[2]
\email{koenecke@cornell.edu}
\affiliation{%
  \institution{Cornell University} 
  \country{Ithaca, NY, USA}
}
\renewcommand{\shortauthors}{Pang et al.}
\newcommand{\draftonly}[1]{#1}
\newcommand{\draftcomment}[1]{\draftonly{#1}}
\newcommand{\rock}[1]{\draftcomment{{\color{cyan}[#1]$_{rock}$}}}
\newcommand{\todo}[1]{\draftcomment{{\color{green}[#1]$_{todo}$}}}
\newcommand{\ak}[1]{\draftcomment{{\color{red}[#1]$_{todo}$}}}
\newcommand{\jc}[1]{\draftcomment{{\color{red}[#1]$_{todo}$}}}

\newcommand{\cc}{consistent conceptualization\xspace}
\newcommand{\labelNum}{28\xspace}
\newcommand{\surveyParticipantNum}{5800\xspace}

\begin{abstract}

Recommendation systems increasingly depend on massive human-labeled datasets; however, the human annotators hired to generate these labels increasingly come from homogeneous backgrounds. This poses an issue when downstream predictive models---based on these labels---are applied globally to a heterogeneous set of users.  We study this disconnect with respect to the labels themselves, asking whether they are ``consistently conceptualized'' across annotators of different demographics. In a case study of video game labels, we conduct a survey on 5,174 gamers, identify a subset of inconsistently conceptualized game labels, perform causal analyses, and suggest both cultural and linguistic reasons for cross-country differences in label annotation. We further demonstrate that predictive models of game annotations perform better on global train sets as opposed to homogeneous (single-country) train sets. Finally, we provide a generalizable framework for practitioners to audit their own data annotation processes for consistent label conceptualization, and encourage practitioners to consider global inclusivity in recommendation systems starting from the early stages of annotator recruitment and data-labeling.
\end{abstract}

\ccsdesc[500]{Social and professional topics~Cultural characteristics\CSLINE}
\ccsdesc[500]{Human-centered computing~Social recommendation}

\keywords{label bias, human annotations, cultural bias, linguistic bias, poststratification, causal inference, recommendation systems}


\maketitle

\input{01_introduction.tex}

\input{02_related_work.tex}

\input{03_case_study.tex}
\input{04_framework.tex}
\input{05_discussion.tex}
\vspace{-0.25cm}
\input{06_conclusion.tex}



\bibliographystyle{ACM-Reference-Format}
\bibliography{bibliography}
\input{99_appendix.tex}

\end{document}

%% file: 01_introduction.tex
\section{Introduction}\label{sec:introduction}
 Labels are used for genre construction and assignment across modes of entertainment like movies, music, and games. Labeling is similarly the goal of many prediction tasks across domains, including computer vision (e.g., art classification~\cite{Lecoutre2017RecognizingAS} and scene understanding~\cite{Patterson2014TheSA}), speech (e.g., accent classification~\cite{Faria2005AccentCF} and sound event annotation~\cite{kim2021}), and natural language processing (e.g., identifying personal attacks~\cite{Wulczyn2016ExMP}, emotionally manipulative language~\cite{Huffaker2020CrowdsourcedDO}, and misinformation~\cite{Mitra2015CREDBANKAL}).

 In order to generate predictions of labels, recommendation systems are often built on extensive datasets labeled by human annotators, which are in turn used to train and evaluate these predictive machine learning (ML) models. To curate a massive labeled dataset, practitioners commonly leverage crowdsourcing, which distributes data labeling microtasks cheaply and efficiently to online crowdworkers~\cite{Buhrmester2016}. Crowdsourcing is similarly used by entertainment platforms to highlight user-generated labels for shopping or games~\cite{amazon,steam}. While much research has historically been conducted on the variable quality of crowdsourced annotators---for example, on popular crowdsourcing platform Amazon Mechanical Turk~\cite{snow2008cheap,Ipeirotis2010,Kees2017,Peer2013}---only recently have researchers turned to studying the demographic makeup of human annotators.

\input{floats/figure_1.tex}

This newer body of work stresses that an annotator's demographic identity shapes their understanding of the world, leading to different labeling outcomes~\cite{Posch2018CharacterizingTG, Ghosh2021DetectingCB, litw, northcutt2021pervasive}, and that responsible data labeling should consider \emph{who} the labelers are~\cite{diaz2022crowdworksheets}. However, crowdsourcing platforms often lack diversity across annotators. On Amazon Mechanical Turk, over 91\% of the annotators come from either the US or India~\cite{Difallah2018, ross2010crowdworkers}. The broader data-labeling industry increasingly outsources labeling tasks to specific developing countries where hiring annotators costs significantly less~\cite{surgeai, Pak2021, Murgia2019, Hale2019}. 

The resulting labels, as part of massive labeled datasets, are then used to train predictive models that in turn predict new labels that are often applied to a global audience. However, problems arise when the demographic makeup of the annotators (often hailing from only one or two countries) is not representative of the demographic makeup of the end users of the ML-predicted labels. For example, an American company found a 30\% label error rate  when auditing Google's ``GoEmotion'' dataset wherein English language text comments were each labeled with emotions; while Google's annotators were all native English speakers from India, there was a cultural divide over linguistic interpretations of the American-centric English in the text \cite{surgeai}.

With non-representative annotators, predictive models overfit training data generated by a limited number of countries and/or languages. In turn, the same models are then used to generate labels for broader global populations. This global reach often takes the form of recommendation systems~\cite{Pu2012,grouplens}, which take as input human-annotated labels that may carry cultural assumptions, and generate recommendations for users assuming the same cultural assumptions hold. This allows the annotators' cultural biases to percolate through the recommendation system. For example, when Netflix first developed its recommendations of ``genre rows'' using a global algorithm in 2015 \cite{GomezUribe2015}---at which point it had not yet reached substantial viewership in smaller countries---Netflix used recommendations trained on viewers predominantly from the US and developed Western countries~\cite{Gaw2021} to generate recommendations in smaller countries.

We refer to the group-based differences in annotator's labels---e.g., between English speakers from India vs. English speakers from America, or between Netflix watchers in small countries vs. Netflix watchers in large countries---as ``inconsistent conceptualization'' of labels.  When presenting labels to a global audience, whether the labels reflect a movie genre or a game tag, it is ideal to ensure that labels are ``consistently conceptualized'' across countries and languages to minimize user confusion. Additionally, consistent conceptualization is useful to ensure consistent interpretation of predicted label outputs from a recommendation system across different countries and languages. Meanwhile, a label that is \emph{not} consistently conceptualized is not necessarily unusable; however, it may require additional effort to interpret and correct for appropriate country- or language-biased label predictions.  Evaluating a label for consistent conceptualization allows stakeholders to explain and/or mitigate discrepancies in global user behavior.

Given the high impact of global-scale recommendations, our research question is: \textbf{how can we audit label annotations for better generalization to a global audience}? Per \autoref{fig:fig_1}, our paper's focus is on auditing the annotation process for bias---before labels are input to any predictive ML models such as recommendation systems.

In this paper, we examine the upstream labeling process by focusing on a case study of video game labels annotated by 5,174 survey participants spanning 16 countries and 9 languages.  We first audit whether game labels are conceptualized consistently across demographics, and explore why inconsistent conceptualization may occur---due to cultural and translation-based phenomena. We then quantify the bias that would arise from using homogeneous annotators from only one country, and find that models predicting label annotations perform better when they are trained on data labeled by heterogeneous annotators---especially so for inconsistently conceptualized game labels. Finally, we provide a framework for practitioners to follow when auditing analogous labels and suggest methods to mitigate bias across demographics.

%% file: floats/figure_1.tex
\begin{figure*}
    \centering
    \includegraphics[width=0.7\textwidth]{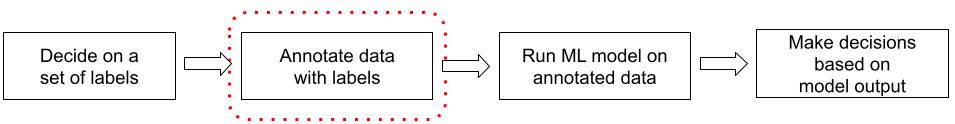}
    \caption{A pipeline of four stages at which labels and machine learning algorithms (such as recommendation systems) can perpetuate bias. This paper focuses on the second stage, auditing label annotations.}
    \label{fig:fig_1}
\end{figure*}

%% file: 02_related_work.tex
\section{Related Work}

As recommendation predictions increasingly influence numerous aspects of our daily lives~\cite{mehrabi2021survey}, more concerns have been raised about the robustness of human-annotated datasets~\cite{Aroyo2015TruthIA, parmar-etal-2023-dont}, and more research has been conducted on their potential to reinforce bias in society~\cite{zou2018ai}, which can arise at any of the four stages in the ML pipeline illustrated in \autoref{fig:fig_1}.
In this section, we review the prior work on measuring bias in machine learning following \autoref{fig:fig_1}, focusing on challenges specific to game genre labels and recommendation systems.

First, deciding on the labels themselves is a difficult task prone to bias. A preliminary question is who generates the labels; the answer could range from the platform themselves, to users on a platform, to an external board.  For any of these parties, the following concerns must be considered; we focus on game labels as a running example.
\begin{enumerate}
\item Generating overly broad labels (such as Action-Adventure, which is so wide-ranging that it becomes less meaningful)~\cite{clarke2017video}
\item Generating overlapping labels such that an annotator could over-apply labels to a game---such as tagging a single game with labels for Action, Adventure, and First-Person Action, hence confusing users and making game marketing more difficult~\cite{clarke2017video}
\item Being overly defined by social conventions, so that understanding labels is challenging for users unfamiliar with the domain or from different cultures~\cite{lee2005challenges}
\item Changing label definitions due to ``genre colonization,'' occurring when elements of one label are used to rationalize and legitimize changes in another label over time~\cite{beghtol2005ethical}
\item Crowdsourcing label generation, yielding noisy data that represents unrelated aspects to the desired labels~\cite{clarke2017video}
\item Choosing relevant levels of a label (e.g., ``Is this an action game?'' versus ``On a scale of 1-5, how much do you believe this is an action game?''), usually through absolute rating scales, often in the form of a binary or Likert scale~\cite{Kittross1959TheMO}. Divergent interpretations of a scale based on abstract text descriptions can become a source of disagreement and inconsistency across annotators~\cite{Weijters2016TheCS}.
\end{enumerate}

Second, annotating the data using the labels determined in the first stage is a non-trivial task that includes recruiting annotators, which is prone to bias along with the annotators' potentially inconsistent conceptualization of labels. Prior work has been conducted to account for the demographic makeup of annotators. Barbosa and Chen~\cite{barbosa_rehumanized} created a crowdsourcing framework that allocates microtasks considering crowdworkers' demographics and compensation. Gordon et al.~\cite{jurylearning} contributed a deep learning architecture that models every annotator in the dataset, amplifying minority groups' labels based on the model context through a metaphor of a jury. Our work, while also examining this second piece of the pipeline, differentiates from this earlier work by focusing on the specific interactions between annotator groups and the conceptualization of the label itself arising from cultural and linguistic differences.

Third, there may be bias in the model itself. This step involves bias mitigation strategies including pre-, in-, and post-processing. Pre-processing transforms training data to reduce potential sources of bias~\cite{dAlessandro2017ConscientiousCA, mehrabi2021survey}; in-processing often modifies the learning objective in the training process~\cite{Berk2017ACF, mehrabi2021survey}; and post-processing takes place after training by accessing a holdout set and transforming model scores~\cite{dAlessandro2017ConscientiousCA, mehrabi2021survey}. 

Fourth, there may be bias in the decision-making that arises as a result of model outputs. For example, an ML model may output predictions for whether different games are classified as 'action' games or not. A recommendation system might then take those predictions as input, and decide how to rank them in order to recommend games to users~\cite{Schafer2007CollaborativeFR}. Much progress has been made in measuring the ranking effectiveness of recommender systems on diverse groups~\cite{ekstrand2018all, mehrotra2017auditing}. Numerous related bias mitigation techniques have been proposed specifically for recommendation systems~\cite{ziweifighting, zhu2020unbiased, burke2017multisided, himan2019multi, chen2020bias, yunqi2021Tutorial}. 

This paper focuses on the second stage of the pipeline in \autoref{fig:fig_1}: auditing the data annotation process. However, the other three pieces of the pipeline are similarly critical to evaluating bias, and any biases arising from the first stage of the pipeline would necessarily confound the second stage. As such, a critical assumption of the case study presented in this work is that our game labels from the first stage were appropriately conceived to minimize the concerns itemized above.

%% file: 03_case_study.tex
\section{Case Study: Video Game Labels}

To illustrate how we examine the \cc of labels, we present a case study of video game labels. Game labels are not only pertinent for understanding the landscape of the video game industry (forecasted to grow by 50\% over the next five years~\cite{christofferson_james_obrien_rowland_2022}), but also lead to better user experiences on gaming platforms~\cite{gameopedia_2022}. Appropriate game labels allow for more direct searches, and can feed into more accurate recommendation systems for users.

We conducted a large-scale survey in the Xbox gaming ecosystem wherein respondents annotated different games using given labels (e.g., ``Is [game] an action game?''). Our analysis reveals that we can identify inconsistently conceptualized labels across countries, and we use one such label as a running example throughout the case study. We then present two explanations for why certain labels are inconsistently conceptualized, based on (1) cultural differences and (2) translation differences. Finally, we show that label predictions can be improved if trained on a dataset of a heterogeneous global population of annotators rather than a small homogeneous population from a single country.

\subsection{Data}\label{sec:data}

Our analysis is based on an email survey conducted in May-June 2022 targeting Xbox gamers.
The survey was run in 16 countries: Argentina, Chile, Colombia, Mexico, Brazil, Germany, Greece, Japan, Poland, Saudi Arabia, South Korea, India, Nigeria, Singapore, South Africa, and the United States. These countries were selected to provide coverage of all distinct regions in the Inglehart–Welzel Cultural Map~\cite{inglehart2000world}.

The survey first allowed the respondent to select which of 11 popular video games they had played in the past year. Then, for the games the respondent selected, they were asked to annotate the game using any of \labelNum labels presented (such as ``action'', ``cozy'', etc.). We also collected respondent demographic information such as age, gender, gaming frequency, and gaming device.

For respondents in countries where English is not an official language, the entire survey text was translated to their local language.\footnote{The non-English survey languages were: Spanish, Portuguese, German, Greek, Japanese, Polish, Arabic, and Korean.} In this process, we performed translation and back-translation by native speakers to ensure translation consistency across languages~\cite{brislin1970back}. We additionally recruited a smaller sample of global Xbox Ambassadors~\cite{xboxamb}---individuals who are known to be bilingual and opt-in to additional communications from Xbox---to respond to the survey in English; the Xbox Ambassadors were primarily based in the US, Brazil, German, Mexico, and Poland. 
To summarize, all individuals with IP addresses in India, Nigeria, Singapore, South Africa, and the US were served only surveys in English; individuals in all other countries were served surveys either in their local language or---if they are an Xbox Ambassador known to be bilingual---English.

\input{floats/figure_2.tex}

In total, we received 5,174 survey responses with complete demographic information; the dominant respondent demographic was US-based men in the 25-44 age range, which is reflective of both the demographics of individuals who opt in to receive emails from Xbox, and the selection bias arising from individuals who choose to proactively fill out the survey for no reward. We explore and statistically correct for the respondents' demographic skew relative to overall gamer populations in ~\autoref{sec:eval}. Full details of our survey, including respondent demographics and survey question phrasing, can be found in \autoref{appendix:survey}.

\subsection{Evaluating Labels for Consistent Conceptualization} \label{sec:eval}

We begin by examining game labels that are inconsistently conceptualized across different demographic groups, and to what extent. To do so, we perform multilevel regression and post-stratification (MRP)~\cite{little1993post,wang2015forecasting} to address the imbalanced nature of our dataset. We first run a Bayesian logistic regression using survey respondents' demographic attributes as covariates to generate a distribution of predictions for whether or not a certain label would be annotated for each game. Then, the ensuing predictions are weighted via post-stratification to resemble the known joint distributions of age, gender, and country demographics for Xbox players writ large.\footnote{Further implementation details can be found in \autoref{sec:appendixmethods}; data is proprietary to Microsoft. Code to reproduce our methods can be found at \url{https://github.com/koenecke/crosscultural_labelconsistency/}.} In doing so, our post-stratified estimates of label selection are more accurate over subgroups~\cite{park2006state}---specifically, upweighting respondents with smaller sample sizes such as female or non-US respondents~\cite{wang2015forecasting,popsupport2023}. Note that for the ensuing analyses, we remove respondents from India and Saudi Arabia, as well as non-binary individuals, due to recruiting only 1 or fewer respondents answering the survey for multiple games; however, we comment on extensions of our model to these subgroups in \autoref{sec:f1scores}. We similarly exclude anonymized Game 11 from our analysis due to low survey response rates detailed in~\autoref{appendix:survey}.

As a running example, in \autoref{fig:fig_2} we show the post-stratified predictions of the ``high replayability'' label for each demographic subgroup regarding a certain popular action game (anonymized in this paper). 
We consider the ``high replayability'' label for this game to be inconsistently conceptualized per the green line in \autoref{fig:fig_2}: our model estimates less than a 39\% probability that gender- and age-representative Korean gamers would consider the game to be highly replayable; in a majority vote among Koreans, this game would not be given the ``high replayability'' label. In contrast, we estimate a 59\% probability that gender- and age-representative American gamers would consider the game to be highly replayable; in a majority vote among Americans, the game would be labeled with the inverse of the Korean consensus.\footnote{Estimated shares in \autoref{fig:fig_2} are aggregated for each x-axis value; for example, the roughly 40\% likelihood of a Korean gamer annotating the game as highly replayable is a weighted average based on the true underlying Xbox population of Korean gamers, based on the proportions in each age, and gender intersection---e.g., weighted by the share of 18-24 year old Korean men, the share of 25-34 year old Korean women, and so on. Similarly, the post-stratified estimate for female gamers is based on a weighted average of of underlying Xbox population data on each country and age breakdown, among women; the post-stratified estimate for gamers aged 18-24 is based on a weighted average of underlying Xbox population data on each country and gender breakdown, among 18-24 year olds.} A majority vote would result in this action game being annotated with the ``high replayability'' label in some countries (Poland, Nigeria, Germany, Colombia, Greece,  Mexico, Brazil, and the US), but not it would not be annotated as such in other countries (Korea, Japan, Singapore, South Africa, Chile, and Argentina). 

In contrast, \autoref{fig:fig_2} also presents an example of a consistently conceptualized label, ``zen'' (red line): all country subgroups---and gender and age subgroups---agree that the action game should not be annotated with this label. We further note that consistent conceptualization of labels is dependent on game. For example, the ``zen'' label is more likely to be consistently conceptualized (and not be annotated) for an action game than, say, a cozy game. As such, we consider inconsistently and consistently conceptualized labels on the basis of game-label pairs (of which there are 280 in total). We recognize that deciding whether a label is inconsistently conceptualized is inherently a source of bias, given that the practitioner must decide which subgroups to examine for inconsistency, and what threshold cutoffs are relevant. 

While we cannot provide a one-size-fits-all formal mathematical definition for inconsistent conceptualization for all domains, we use this terminology to describe labels that have high variance across and within demographic subgroups, and when the subgroup mean straddles the 50\% mark for binary labels. For the ensuing analyses in this paper, we categorize 55 game-label pairs as ``inconsistently conceptualized'' labels, having cross-country standard deviations greater than 0.05 among post-stratified estimates (calculated from the ``Country'' facet of \autoref{fig:fig_2}) and with post-stratified country-level estimates both over and under the majority 50\% line. Cross-country standard deviations for our surveyed labels and anonymized games are presented in \autoref{fig:fig_tile} in \autoref{appendix:heatmap_country}.

Our designation of inconsistent labels is robust across reasonable definitions. The post-stratified estimates presented in \autoref{fig:fig_2} are similar to the unweighted raw survey results (\autoref{fig:appendix_prop} in \autoref{app:survey_figs}), indicating that our survey respondent pool is reflective of the true underlying gamer demographic distribution from which we sampled.\footnote{Our analyses are robust to choice of standard deviation size and threshold-crossing; results hold directionally (though in smaller magnitudes) even when 132 (nearly half of) game-label pairs are categorized as ``inconsistently conceptualized.'' We further confirm that our 55 ``inconsistently conceptualized'' labels are reasonable by calculating the Spearman correlations of the \emph{rankings} of post-stratified estimates from \autoref{fig:fig_2}. For example, for a certain action game, we can rank the 28 labels in order of most-annotated to least-annotated for the game among one country's annotators (e.g., ``action'' has rank 1, ``zen'' has rank 28). We can then compare those rankings to the rankings elicited from another country's annotators. We find that while the overall ranking similarity is quite high between country pairs (Spearman correlation seeing a minimum of 0.7 and mode of 0.97, histogram provided in \autoref{sec:appendixmethods}), the inconsistently conceptualized labels are more prone to be ranked out-of-order between country pairs: the median rank-difference among inconsistently conceptualized labels is 1.5, whereas the median rank-difference among consistently conceptualized labels was only 1.0.\label{rank_footnote}}

Our case study focuses on the inconsistent conceptualizations across countries, where there tends to be higher variance across game labels relative to age- or gender-groups of label annotations. For example, in the ``high replayability'' example in \autoref{fig:fig_2}, we see a 20-percentage-point spread of estimates across countries versus a roughly 10-percentage-point spread of estimates by gender. We now turn to understanding why country-level differences arise.


\subsection{Explanations for Inconsistent Conceptualization}

We present two explanations for the country-level differences in label conceptualization across our surveyed labels. The first invokes cultural frameworks from the sociology literature, and the second involves semantic differences in language translations.

\input{floats/figure_3_4}

\subsubsection{\textbf{Cultural Distances}}

Much prior work has been conducted across sociology to measure cross-cultural similarities via indices, including the World Value Survey~\cite{inglehart2000world} used for our country selection process per Section \ref{sec:data}, and the Hofstede Model of Culture~\cite{hofstede1980culture}. The Hofstede Model of Culture consists of six dimensions of national cultures (developed over the course of 1991~\cite{hofstede1980culture} to 2010~\cite{hofstede2010long, bond1991beyond}): power distance, uncertainty avoidance, individualism/collectivism, masculinity/feminity, long/short-term orientation, and indulgence/restraints. Each country is mapped to six numeric indices ranging from 0 to 100\%.

First, we generate a Cultural Distance Index based on the subset of Hofstede dimensions that we hypothesize can explain different countries' label annotations. To determine which dimensions are relevant, we run a logistic regression similar to~\autoref{sec:eval} across all survey responses on all games and labels, but replace the country categorical covariate with the six Hofstede dimensions. We find that the coefficients on two dimensions are significant at the p < 0.001 level and have the highest magnitude of all dimensions; these two dimensions are ``uncertainty'' and ``orientation'' (regression results are presented in ~\autoref{tab:hofstedereg} in ~\autoref{sec:appendixmethods}).\footnote{A third Hofstede dimension, ``individualism,'' is weakly significant (p < 0.1) with small coefficient magnitude in the regression specification. Individualism/Collectivism refers to the ``integration of individuals into primary groups,'' where individualism aligns with cultures wherein ``ties between individuals are loose: everyone is expected to look after him/herself and his/her immediate family'' ~\cite{hofstede1980culture}. Our analyses are robust to including the individualism dimension in our cultural distance index, though the relationship plotted in \autoref{Fig:fig_3} is very slightly weaker when including ``individualism'', still yielding a Spearman correlation of -0.4.}

The \emph{uncertainty} Hofstede dimension is based on the concept of ``uncertainty avoidance,'' defined as ``the extent to which the members of a culture feel threatened by ambiguous or unknown situations''~\cite{hofstede1980culture} under the premise that ``extreme ambiguity creates intolerable anxiety.'' Hofstede's \emph{uncertainty} dimension measures how much or little a society copes with anxiety by minimizing uncertainty (a high Hofstede uncertainty dimension indicates low tolerance for uncertainty). This is considered a proxy for the threat of change to a culture. 
Meanwhile, the \emph{orientation} Hofstede dimension is on the spectrum between short-term and long-term orientation. Short-term orientation is defined as ``the fostering of virtues related to the past and present---in particular, respect for tradition, preservation of `face,' and fulfilling social obligations''; meanwhile, long-term orientation is defined as ``the fostering of virtues oriented towards future rewards---in particular, perseverance and thrift.''~\cite{hofstede1980culture}. Both uncertainty and orientation dimensions are highly relevant to the labels with which one might annotate video games, wherein ambiguity, surprise, patience, and reward-seeking behaviors are often key components~\cite{lukosch2017gender, kuo2022exploring}.

We hence calculate a Cultural Distance Index (CDI) defined as the Euclidean distance between country pairs in the uncertainty and orientation dimensions; for example, Argentina and Mexico have a very close CDI of 0.06, whereas Korea and Nigeria have a nearly polar opposite CDI of 0.92. We then explore whether countries that are more culturally similar---i.e., have a smaller CDI---also annotate game labels more similarly. 

To determine game label annotation similarity by country pairs, we calculate the Pearson correlation between each country's post-stratified estimates across games and labels. Specifically, for each game and label combination, we calculate one post-stratified estimate of each country's likelihood of annotating the game with that label (e.g., roughly 60\% for the US, for a specific action game with the ``high replayability'' label, per~\autoref{fig:fig_2}). We then compare estimates for ``inconsistently'' and ``consistently'' conceptualized labels in aggregate. Hence, for each country, we generate one 55-length vector of ``inconsistently conceptualized'' game-label annotation estimates, and another vector of the remaining 225 ``consistently conceptualized'' game-label annotation estimates. For each pair of countries, we calculate the Pearson correlation between their two 55-length ``inconsistently conceptualized'' vectors, and the Pearson correlation between their two 225-length ``consistently conceptualized'' vectors. The relationship between CDI and label similarity, stratified by label consistency, is plotted in \autoref{Fig:fig_3}. 

The first takeaway is a confirmation of our definition of ``inconsistently conceptualized'' labels in \autoref{sec:eval}: indeed, the game-label pairs we deem to be consistently conceptualized yield high (often close to perfect) annotation correlation regardless of country pair. Meanwhile, inconsistently conceptualized labels have much lower between-country agreement. The second takeaway is that as cultural distance (CDI) increases---that is, cultural similarity decreases---game label annotation similarity also decreases. While this is a weak decreasing effect among consistently conceptualized labels, we in contrast see a stronger negative relationship (slope -0.3, with a weakly negative Spearman correlation of -0.4) for inconsistently conceptualized labels. For example, the culturally close countries of Mexico and Argentina have a perfect annotation correlation of 1.0 for consistently conceptualized labels, and lower but still high correlation of 0.8 for inconsistently conceptualized labels. Meanwhile, the culturally distant countries of Korea and Nigeria have a high annotation correlation of 0.9 for consistently conceptualized labels and the lowest (0.0) annotation correlation for inconsistently conceptualized labels.

To summarize, we find that the lower the CDI (i.e., the more ``culturally similar'' a pair of countries are per the Hofstede model), the more similarly respondents from those countries tend to annotate game labels. While we cannot draw concrete causal conclusions from specific cultural differences to general game-label conceptualization consistency, we do find that culture may play a non-negligible role in understanding country-based differences in game label annotations. 

\subsubsection{\textbf{Lost in Translation}}

In addition to modeling culture, another explanation for inconsistently conceptualized labels across countries lies in the label text itself. Translation to a different language can lead to the loss of semantic precision~\cite{peskov-etal-2021-adapting-entities, santy-language}, even if professionally translated, due to the fundamental difference in cultural conceptualization of a word.  To determine whether language itself has an impact on game label annotations within a country, we compare respondents surveyed in English to a demographically-similar set of respondents surveyed in their local language, using Mahalanobis distance matching~\cite{mahalanobis1936generalised,ROSENBAUM1983,caliendo2008some}. To ensure we capture differences in semantics and not differences in player experience, we subset to respondents who play games in English ``frequently'' or even more often. We then match the Xbox ambassadors (surveyed in English) to their compatriots (surveyed in their local language) on dimensions of age, gender, and gaming experience.\footnote{There may, of course, be additional confounders---e.g., the Xbox Ambassadors could be a fundamentally different survey population than the other survey respondents responding to the email survey---however, these are the extent of the personal demographics we capture in our survey.} Covariate balance was confirmed between the English and local survey language respondents (balance plots are reported in \autoref{fig:balplots} in \autoref{sec:appendixmethods}). We then estimate the effect of survey language on label annotation using a logistic regression on the matched sample, estimating standard error with cluster-robust variance.

Overall, we find that the share of significant survey language effects is slightly higher among inconsistent labels (36\%) relative to consistent labels (34\%). To showcase the variability in language effects, we turn to another inconsistent label as an example: ``makes me feel amused or laugh'' for a specific action game.\footnote{Our running label examples of ``high replayability'' and ``zen'' from \autoref{fig:fig_2} yields extremely small matched counts ($n<4$ respondents for each language) for the specific action game; here, we use a different action game for the ``makes me feel amused or laugh'' example. For the sake of completion, we note that the inconsistently conceptualized ``high replayability'' label sees an increase in English relative to Polish (statistically significant at the 1\% level), but no statistically significant language effects in Brazil or Mexico. Meanwhile, the consistently conceptualized label ``zen'' sees a 17 percentage point decrease in English relative to Portuguese (statistically significant at the 1\% level), but no statistically significant language effects in Mexico or Poland. There were not enough matched participants in Germany for this particular game to comment on the English-German effect.} 
In \autoref{Fig:fig_4}, we present the grouped mean differences for individuals annotating a certain action game as such.
For demographically matched respondents in Brazil, those surveyed in English had a 21 percentage point \emph{lower} respondent share answering that the game made them ``feel amused or laugh,'' relative to being surveyed in Portuguese. A similar decrease, though with a larger magnitude and statistically significant, was observed between Polish respondents answering in English (63 percentage points lower) relative to Polish.
Meanwhile, for demographically matched respondents in Germany, those surveyed in English had a statistically significant 38 percentage point \emph{higher} respondent share agreeing with the label, relative to being surveyed in German. A similar increase, though not statistically significant, was observed among Mexican respondents answering in English (22 percentage points higher) relative to Spanish. Full statistics, including confidence intervals and p-values, are reported in ~\autoref{tab:bilingual} in \autoref{sec:appendixmethods}.

We make several notes on the directions of English language effects on bilingual respondents. 
First, the English effect is independent across countries; for a specific label for a specific game, there is no reason to expect that \emph{all} languages' translations of that label would consistently yield either all increases, or all decreases, in the probability of annotation relative to English.
Second, there is no reason that the direction of the English effect must persist across labels for a local language (e.g., there may be some labels that lose meaning when translated to that language, but other labels may have no translational difference; furthermore, some label translations may yield a higher probability of label annotation, whereas other label translations may yield a lower probability of label annotation).
Third, language-based differences in a label need not persist across games (e.g., if a label---even conceptualized differently---is clearly not relevant to the game being annotated, the language difference may not be as strong).

Overall, we find that differences in label \emph{translation} do contribute in varying degrees to the inconsistent conceptualization of labels across countries, but are highly dependent on both the label itself and the language (as well as, of course, the quality of translation).

\subsection{Heterogeneous annotators improve global game label predictions}\label{sec:f1scores}

We have established that some video game labels are inconsistently conceptualized, and provided two potential sociological reasons for this to be the case: cultural differences based on Hofstede dimensions, and text translation biases. While this is not in and of itself a concern, extrapolating game labels based on only one culture's annotations to a broader global population compounds labeling biases. We demonstrate that incorporating diverse participants across different countries can lead to a more accurate and representative label decision for a global audience.

Per Section \ref{sec:introduction}, it is often the case that a homogeneous group of annotators are hired to label data (e.g., from a certain country)~\cite{surgeai, Pak2021, Murgia2019, Hale2019}. Another tactic for eliciting annotators is by using an expert focus group, though this can fall prey to a similar homogeneity concern. When this homogeneous group's annotations are used as the basis of recommendation systems applied globally, the resulting recommendations for non-annotator countries may not align with those countries' conceptualizations of the same labels.  For example, if a game is annotated as a ``highly replayable'' game based on a US-centric understanding of what this label means, it may lead to user confusion when the game is recommended to a gamer in Korea, who would not consider the same game to be ``highly replayable.''

We show this in aggregate by simulating label predictions on different country subgroups, for ``inconsistently conceptualized'' labels that are especially difficult to predict with high accuracy. We begin by generating two different train sets: first, we generate a dataset based only on survey responses from one homogeneous group's game label annotations---American gamers. Second, we generate a new train set based on survey responses from a globally representative population excluding American gamers. 
Then, we train the same logistic regression model on each of these two train sets, using covariates including game, label, and respondent demographics. This yields two predictive models, one trained on a ``homogeneous'' annotator group and the other trained on a ``heterogeneous'' annotator group. We then compare the two models' performance on a held-out test set (representing 30\% of the collected data) on globally representative survey respondents' game label annotations.\footnote{Our results are robust to sampling the homogeneous and heterogeneous train set in different ways per \autoref{fig:appendix_figure7_representative}, e.g. demographically representative stratified sampling, and oversampling (so the homogeneous train set size matches the heterogeneous train set size).}

\input{floats/figure_5.tex}

Our results are shown in \autoref{fig:fig_5}; at a country level, we aggregate F1 scores on the held-out test set across both inconsistently conceptualized labels and consistently conceptualized labels. First, we confirm that consistently conceptualized labels are relatively easy to predict (yielding a similarly high F1 score across countries in the test set) regardless of the country distribution underlying the training data: as we showed in \autoref{Fig:fig_3}, most countries annotate these labels quite similarly. Second, we see improvement in F1 scores across all countries---including for American gamers---when the heterogeneous (global) dataset is used to train the model rather than only the homogeneous (US-based) train set. Across the entire test set, we see a 7.7\% improvement in F1 score; further results are presented in \autoref{sec:appendixmethods}. More granularly, recall that we found Korea and Nigeria to be dissimilar countries (both along our Hofstede-based Cultural Distance Index and based on label annotation correlations). Continuing our running example from \autoref{fig:fig_2}, while the homogeneously-trained model (trained on gamer demographics) predicts similar ``highly replayable'' label annotation shares of 55.47\% for Korea and 55.98\% for Nigeria, the heterogeneously-trained model (additionally trained on a \emph{country} covariate) predicts a much more distinct set of annotation shares: 38.85\% for Korea and 51.95\% for Nigeria, which is far more reflective of the wide difference in true label annotations per \autoref{fig:fig_2}.

Furthermore, using a globally diverse training set allows for better out-of-sample predictions. Recall that we excluded responses from India and Saudi Arabia in all prior analyses due to extremely low survey response rates for many games. When including these countries in our test set (despite not being in our train set), we find that the model trained on heterogeneous data yields better prediction accuracy on these countries relative to the model trained on homogeneous data (with improvements ranging from a 2-3\% increase in F1 score). Overall, our results support the intuitive finding that diversifying the training set improves predictions on a diverse test set~\cite{Kim_2019_CVPR,Koh2021,Mitchell2021,buolamwini2018gender,racial_disparity}, and we encourage stakeholders to invest in diversifying annotator populations hired for data labels. 

%% file: floats/figure_2.tex
\begin{figure*}
    \centering
    \includegraphics[width=0.9\textwidth]{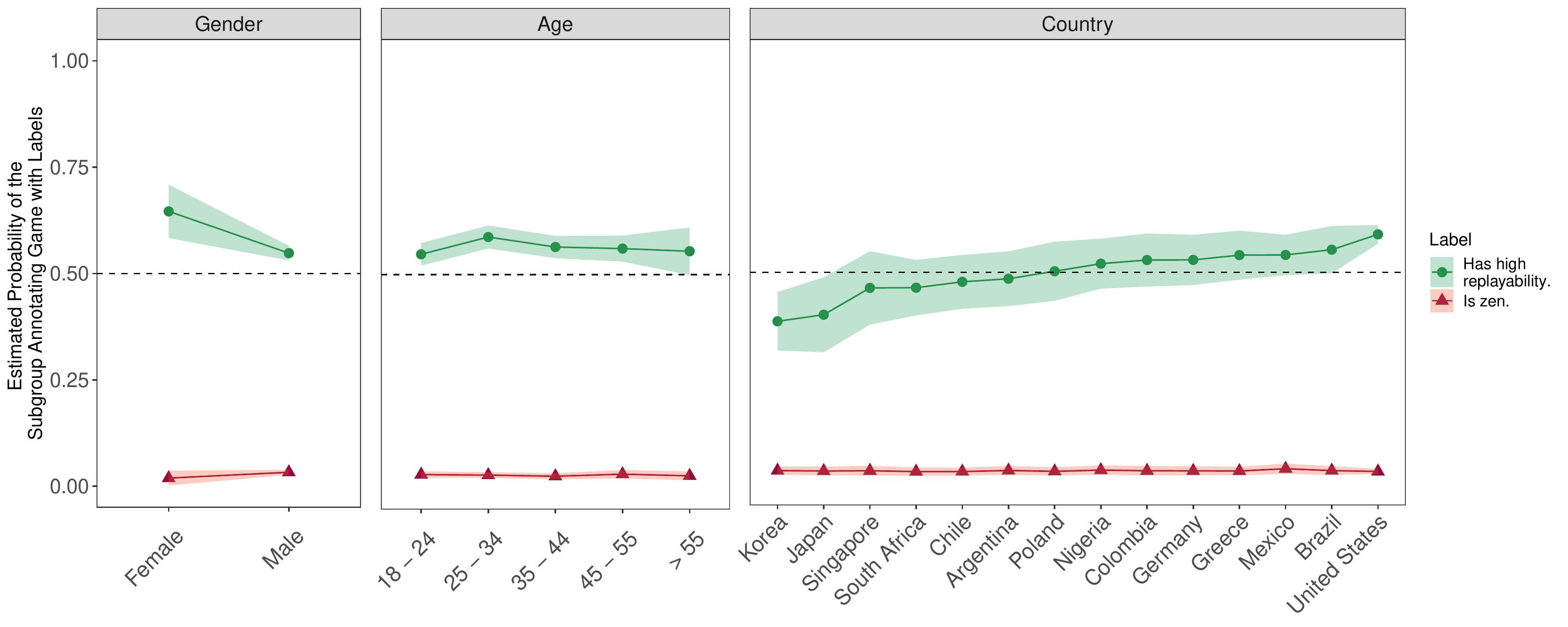}
    \caption{Post-stratified estimates that a certain action game would be annotated with the label ``high replayability'' (green line) or ``zen'' (red line) for subgroups by gender, age, and country. 
    The estimated probability that gamers in Korea would annotate this game as highly replayable is less than 40\%; in contrast, we estimate a 60\% probability that gamers in the United States would annotate this game as highly replayable. The ``high replayability'' label exemplifies what we refer to as an ``inconsistently conceptualized'' label: there is a wide spread of annotation estimates across countries (for this particular pair of ``action'' game), and a majority vote on whether the label applies to a game yields different binary decisions in different countries. In contrast, all demographic subgroups across gender, age, and country, broadly agree that the same game would not be annotated with the label ``zen,'' which we refer to as a ``consistently conceptualized'' label.
    }
    \label{fig:fig_2}
\end{figure*}

%% file: floats/figure_3_4.tex
\begin{figure*}[ht]
\begin{minipage}[b]{0.49\linewidth}
\centering
\includegraphics[width=\textwidth]{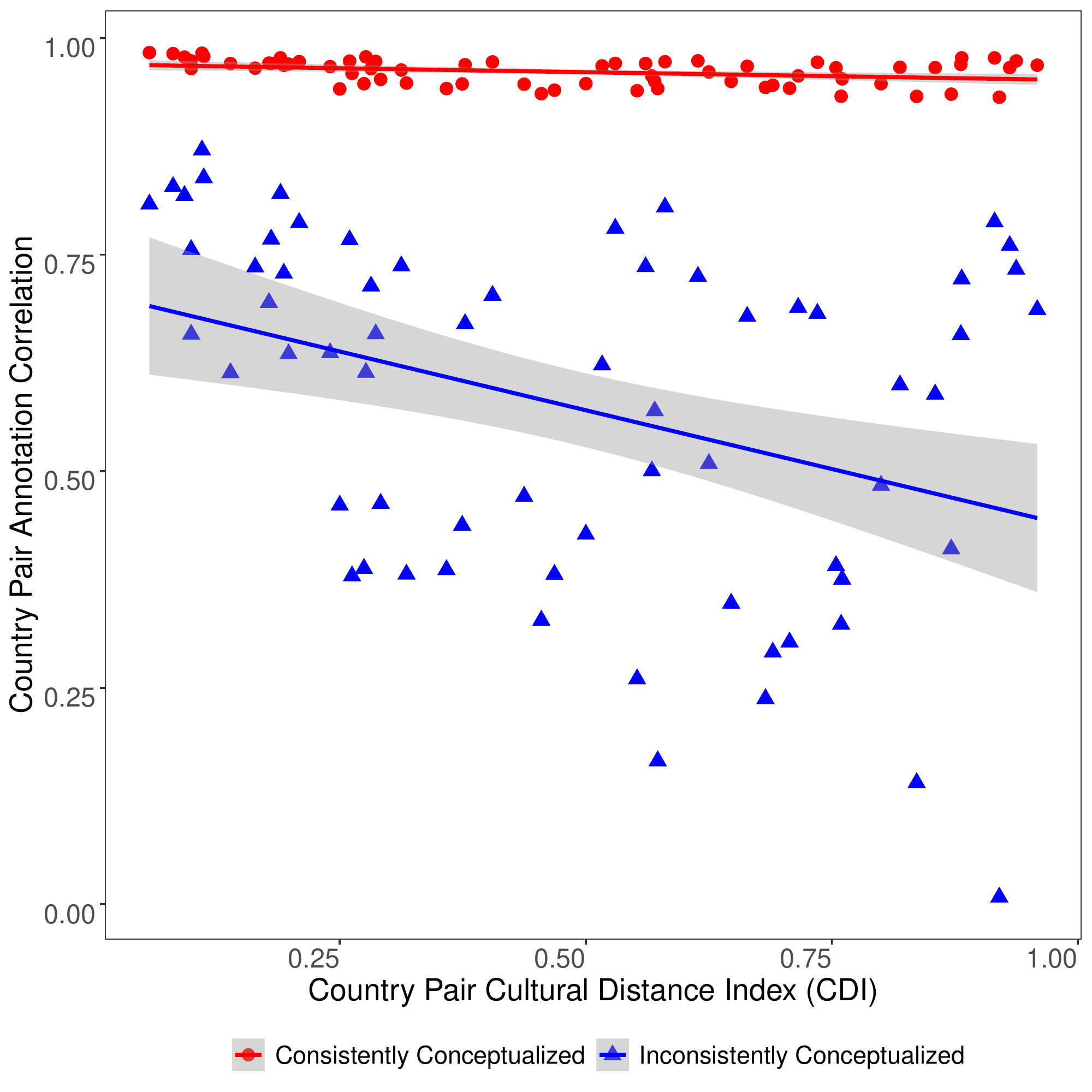}
     \caption{Countries that are more culturally similar to each other (i.e., with lower CDI) also annotate games with more similar labels. There is a negative relationship between a pair of countries' cultural distance index (along the x-axis, greater cultural distance implies countries that are more different---as measured by Hofstede's cultural dimensions of ``long term orientation'' and ``uncertainty'') and the same pair of countries' game-label annotation similarity (along the y-axis)---but only for inconsistently conceptualized labels across games. Meanwhile, we confirm that consistently conceptualized labels do not see differences in annotation similarity regardless of culture, corroborating our definition of consistent conceptualization in~\autoref{sec:eval} (along the y-axis) for pairs of countries. }
     \label{Fig:fig_3}
\end{minipage}%
\hspace{0.01\linewidth}
\begin{minipage}[b]{0.49\linewidth}
\centering
\includegraphics[width=\textwidth]{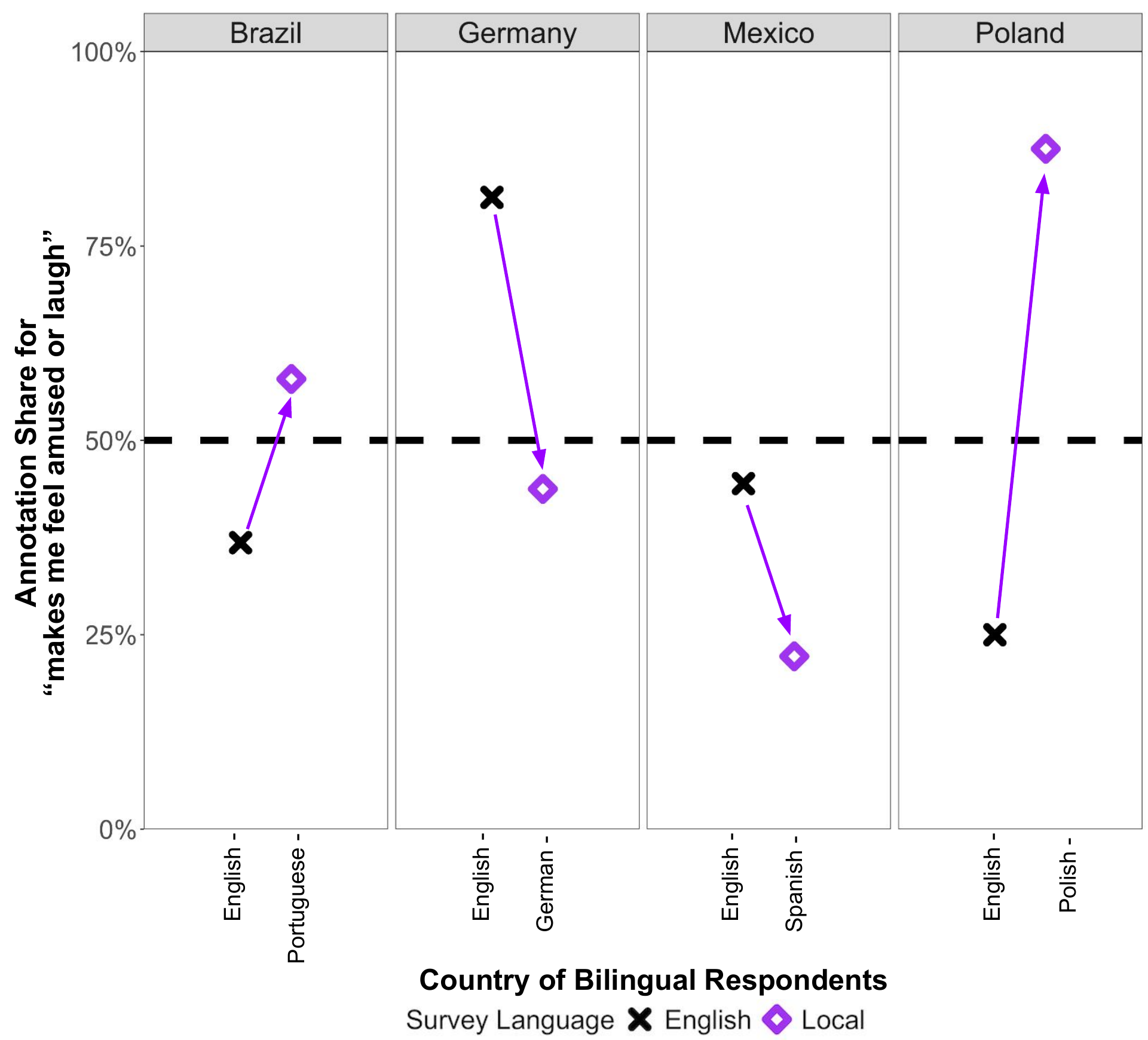}
    \caption{
    For an action game's inconsistently conceptualized label (``makes me feel amused or laugh''), we find that translation differences may contribute to the differences across estimated annotation rates by country. Bilingual respondents in Brazil, Germany, Mexico, and Poland were randomly surveyed in either English or their respective local languages. Within each country, respondents in each survey language were matched on demographic attributes including age, gender, gaming experience, and English gaming experience (matched $n =$ 34, 26, 16, and 16 in Brazil, Germany, Mexico, and Poland for this specific video game).  We plot the matched groups' average annotation shares for whether the action game ``makes me feel amused or laugh''; the majority vote is highly dependent on survey language, indicating a language-based disparity in conceptualization of the label. The difference in share of gamers who would positively annotate the game with this label is statistically significant for two pairs of translations (between English and German, and English and Polish).
    }
    \label{Fig:fig_4}
    \end{minipage}
\end{figure*}

%% file: floats/figure_5.tex
\begin{figure}[t]
    \centering
    \includegraphics[width=\columnwidth]{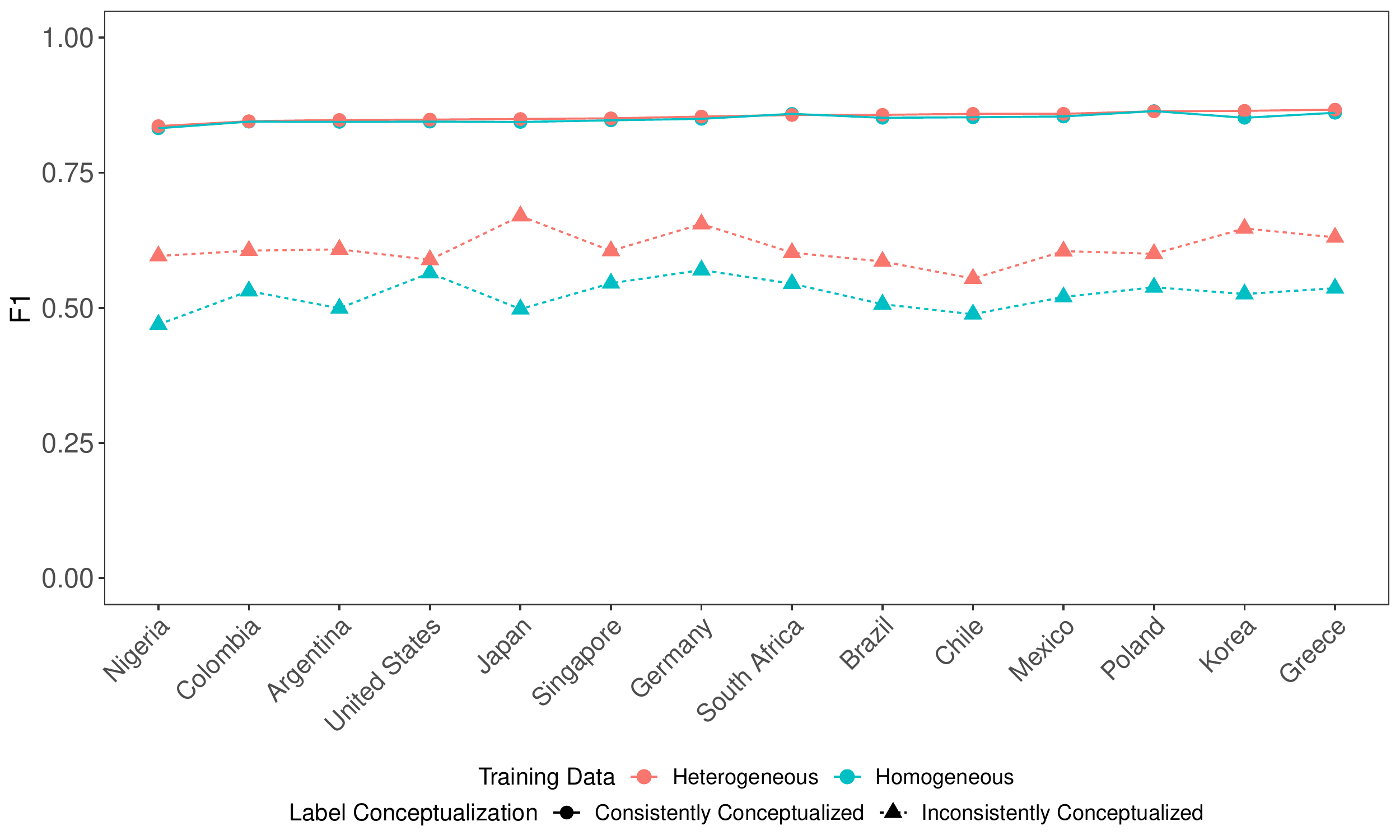}
    \caption{An evaluation of F1 scores across 14 countries from logistic regression models trained on either homogeneous or heterogeneous datasets; model improvement is especially clear for the heterogeneous-trained model on inconsistent labels. The top two overlapping lines represent the high F1 scores achieved by both homogeneously- and heterogeneously- trained models for consistently conceptualized labels. The lower blue triangle line indicates the low F1 scores for a (US only) homogeneously-trained model on a global test set of inconsistently conceptualized labels, which are naturally more difficult to predict. The red triangle line shows improvement in F1 scores for inconsistently conceptualized labels when the model is trained on heterogeneous annotators instead---with improvement even for the US.
    }
    \label{fig:fig_5}
\end{figure}

%% file: 04_framework.tex
\section{Framework: Auditing Data Annotation for Global Inclusivity}\label{sec:framework}

We summarize our research process in a generalized framework for auditing the annotation process. This framework provides a roadmap for practitioners to ask auditing questions on both country and language inclusivity, continuing our running example of game labels. For each question, we further propose actionable next steps for practitioners to take depending on the outcomes of each audit.
Note that our framework should serve as a guide rather than an oracle. While we address country and language questions separately, it is important to additionally consider their interactions in cultural contexts. And, even if labels pass all of the proposed auditing criteria, this does not guarantee the same interpretation across cultures; rather, it simply satisfies our definition of ``consistent conceptualization'' and means that a predictive model taking the label as input would yield less country- or language-based bias for that label.

\subsection{Country-based Auditing}
\subsubsection{Across countries, how much \textit{external} agreement is there on labels of interest?}\label{sec:countryq1}
\begin{itemize}
\item \textbf{Example}: Do more Americans consider a certain game to be an ``action'' game than their Japanese counterparts?
\item \textbf{Method}: Calculate differences in estimated label annotations for each demographic.  To compare estimates across multiple countries, we suggest conducting post-stratified regression analysis~\cite{wang2015forecasting} to adjust for survey respondents' demographic confounders, as done in \autoref{fig:fig_2}. To elicit the causal country effect between a pair of countries, we suggest performing matching (e.g., with Mahalanobis distance matching~\cite{mahalanobis1936generalised}, the method underlying \autoref{Fig:fig_4}). To elicit similarity between country \emph{rankings} of labels within a game, calculate the Spearman correlation between the ranked post-stratified estimates of label annotations (e.g., to see if a game is \emph{always} ranked as ``action'' more than ``zen'' in all countries), the method underlying \autoref{rank_footnote}.
\item \textbf{Action Item}: Determine whether label estimates should be de-biased for certain populations. For example, assume that your model estimates that 70\% of US gamers would annotate a game with the ``action'' label, only 50\% of Japanese gamers would do the same, and otherwise all labels are ranked in the same order for the game in question. In this case, consider applying a de-biasing correction to recommendation systems so that Japanese gamers are not as frequently recommended this game in the ``action'' category, relative to American gamers.
\end{itemize}
\subsubsection{Within each country, how much \textit{internal} agreement is there on labels of interest?}\label{sec:countryq2}
\begin{itemize}
\item \textbf{Example}: Does the concept of an ``action'' game elicit more agreement among annotators within the US than it does among annotators in Japan?
\item \textbf{Method}: Compare the variance of label outcomes (estimated via \autoref{sec:countryq1}), when grouped by country. Higher variances suggest ``inconsistently conceptualized'' labels; see \autoref{fig:fig_tile}.
\item \textbf{Action Item}: Deep dive into how to approach different country markets. For example, if everyone in one country agrees a game is ``action'' whereas this is not the case in another country, consider auditing for cultural effects (e.g., is there less clarity on the concept of ``action'' in the latter country? See \autoref{sec:countryq3}.) or language effects (e.g., does ``action'' does not translate well to that specific language? See \autoref{sec:langq3}).
\end{itemize}
\subsubsection{If low external or internal agreement is found, is there a research-backed explanation relating to cultural differences?}\label{sec:countryq3}
\begin{itemize}
\item \textbf{Example}: If American and Japanese annotators conceptualize the ``action'' label inconsistently, can this be explained by specific cultural factors?
\item \textbf{Method}: Computationally, one can use Hofstede's cultural indices to calculate the cultural distance between these two countries per \autoref{Fig:fig_3}. However, we emphasize that culture is a highly context-dependent phenomenon~\cite{contexualSonja}. To comprehensively understand the sociological reasons for conceptual differences requires consulting experts in these cultures; reducing country culture to an index will not paint the full picture of why a label might be inconsistently conceptualized. It is important to incorporate additional research when formulating hypotheses, such as cultural differences in game perceptions~\cite{Law2009GenderAC, Zendle2023,Brckner2019ExploringCD, chencute}.
\item \textbf{Action Item}: Consider research across the sociology and Human Computer Interaction literature to understand whether or why different countries may have different conceptualizations of labels. For example, prior work comparing Japanese and American gamers has found differences in preferences regarding replayability and bugs~\cite{Zagal2013CulturalDI}, playtime~\cite{Zendle2023}, and level of control~\cite{Cook2009MethodsTM}.
\end{itemize}

\subsection{Language-based Auditing}
In this section, the questions can and should be benchmarked against the language in which labels were originally conceptualized. We center our questions on English as the default language (because the authors generated labels in English originally) but emphasize that other languages could be used as the benchmark. 
\subsubsection{Across English-speaking countries, how much external agreement is there on labels of interest?}\label{sec:langq1}
\begin{itemize}
\item \textbf{Example}: Is there broad consensus on the ``action'' annotation across India, Nigeria, Singapore, South Africa, and the US?
\item \textbf{Method}: Generate poststratified analyses per \autoref{sec:countryq1} to compare the English-speaking countries within \autoref{fig:fig_2}, or compare pairs of English-speaking countries per \autoref{Fig:fig_3} when matched on demographic confounders~\cite{mahalanobis1936generalised,ROSENBAUM1983}.
\item \textbf{Action Item}: If there is broad consensus across English-speaking countries, it confirms that the cross-cultural differences do not affect the shared conceptualization of a label's meaning in English. If not, focus on country-based differences per \autoref{sec:countryq3}.
\end{itemize}
\subsubsection{For non-English-speaking countries, how much external agreement is there on labels of interest?}\label{sec:langq2}
\begin{itemize}
\item \textbf{Example}: How similar are Japanese annotations of ``action'' games to those of other non-English annotations? How similar are Japanese annotations of ``action'' games to those of English language annotations?
\item \textbf{Method}: Again, we recommend similar poststratified analyses and matching analyses as in \autoref{sec:langq1}, but now comparing specific non-English-speaking countries to all other countries.
\item \textbf{Action Item}: If similar label annotation estimates are found, the labels can be considered consistently conceptualized and language does not appear to play a factor. If not, focus on translation differences per \autoref{sec:langq3}.
\end{itemize}
\vspace{-0.2cm}
\subsubsection{Within each country where English is not an official language, how much internal agreement is there among bilingual speakers annotating English versus non-English labels of interest?}\label{sec:langq3}
\begin{itemize}
\item \textbf{Example}: Is the label ``action'' translated to Japanese in a way that does not convey the same conceptual meaning as in English?
\item \textbf{Method}: Per the methodology in \autoref{Fig:fig_4}, recruit bilingual annotators within a non-English-speaking country (e.g. Japan) and randomize surveys so that labels are annotated by: (a) one set of survey respondents in Engish and (b) another set of survey respondents in Japanese. Then, perform Mahalanobis distance~\cite{mahalanobis1936generalised} or propensity score matching~\cite{ROSENBAUM1983} on respondent demographics (e.g., age, gender, gaming experience, etc.) to elicit the causal difference between the English-language annotations and Japanese-language annotations.
\item \textbf{Action Item}: Within each non-English country with match-ed bilingual survey respondents, if non-significant differences are found between the English and non-English annotations, it is unlikely that label information was lost in translation. It is very likely that findings are label- and language-dependent: e.g., for some labels, the English version may yield more annotations for a game, whereas for others, the English version may yield fewer annotations.
If significant language-based differences are found, we recommend consulting with translation experts and conducting further translation and back-translation. We further recommend understanding underlying cultural differences that could inform linguistic differences per \autoref{sec:countryq3}.
\end{itemize}

%% file: 05_discussion.tex
\section{Discussion \& Limitations}
In this paper, we focus on the data annotation step of the recommendation system pipeline (\autoref{fig:fig_1}), specifically auditing label annotations for global bias. We provide the ``inconsistent conceptualization'' nomenclature to describe the biases discussed, and then illustrate the global biases at hand via a case study on video game labels. By conducting the first large-scale survey of video game labels, we elicit responses from 16 countries in 9 languages and evaluate annotation differences by country and language. 
Our findings from the case study demonstrate that data labels may be inconsistently conceptualized, despite them being uniformly deployed across countries in real-world ML applications. 
Our analyses suggest that inconsistently conceptualized game labels may be explained by both Hofstede's cultural differences and semantic loss through translation. Considering the common practice of hiring annotators from only a single country, we then show that models trained on heterogeneous training data outperform those trained on homogeneous (single-country) training data when applied to global test sets. 
Finally, we provide a generalized framework for practitioners to conduct systematic audits on data annotation---via a series of actionable items---for inconsistent conceptualization of labels across countries and languages, hence fostering global inclusivity across users.

There are several limitations of our work. First, due to the privacy constraints arising from small survey response numbers, our analysis could not encompass some individuals from diverse backgrounds such as non-binary respondents (who were entirely based in the US in our survey). Relatedly, due to legal constraints and privacy concerns, we could not elicit demographic information on race, so intersectionality~\cite{wangIntersectionality, yojinIntersectionality} was not fully captured in our analysis.
Second, while our framework focuses on country and language as key indicators for evaluating global inclusivity in labeling, we acknowledge that using ``country'' as a proxy for ``culture'' in a geographical sense may perpetuate ``cultural hegemony'' in which value systems are shaped and mediated by those in power~\cite{culturalhegemony}. Cultures, even within a country, have changed in dramatic ways over time~\cite{orr2008re}. Hofstede's model, while providing a quantitative and practical approach, may oversimplify the complexity of cultures. 

There are many directions for future work in auditing consistent conceptualization of labels as they percolate through the recommendation pipeline; in conjunction with our proposed framework, we further suggest performing an offline counterfactual evaluation to confirm that using inclusive labels as a feature improves model performance both in aggregate across users, as well as within demographic subgroups. Overall, our findings suggest that inconsistently conceptualized labels arise in human-annotated data, and we encourage practitioners to carefully consider how labels are generated (per the first stage of \autoref{fig:fig_1}) and opt to diversify their annotator groups to yield improved global label predictions downstream. We further suggest that practitioners perform audits of their annotations using the framework provided in \autoref{sec:framework} before making important decisions based on black-box recommendation systems. And, we encourage our findings and suggestions to also be considered outside of recommendation systems, for other predictive models trained on human-labeled data.

%% file: 06_conclusion.tex
\section{Research Ethics \& Social Impact}
This survey study was approved by an institutional IRB and all collected data were anonymized; no PII was collected. We ensured that no analyses were conducted on subgroups of size 5 or fewer to ensure survey respondent privacy.
Our work has broader impact on both the HCI/fairness community and the gaming research community. In the former, we connect the sociology and fairness literature to the data annotation process, a crucial step in the recommendation systems pipeline with massive downstream effects~\cite{green2021data, doc2022hi, winner2017artifacts}. In the latter, we supplement existing theoretical work and small-scale surveys (constrained to specific countries) with larger-scale data analysis of global label conceptualization. We hope that our framework can be applied both in other gaming applications and also across other domains that currently elicit labels from homogeneous annotators to train globally-applied predictive models.

%% file: 99_appendix.tex
\newpage
\onecolumn

\appendix 

\section{Survey Details}\label{appendix:survey}

\subsection{Survey Resondent Demographics}

\begin{figure*}[ht]
\centering
   \centering
   \includegraphics[width=\textwidth]{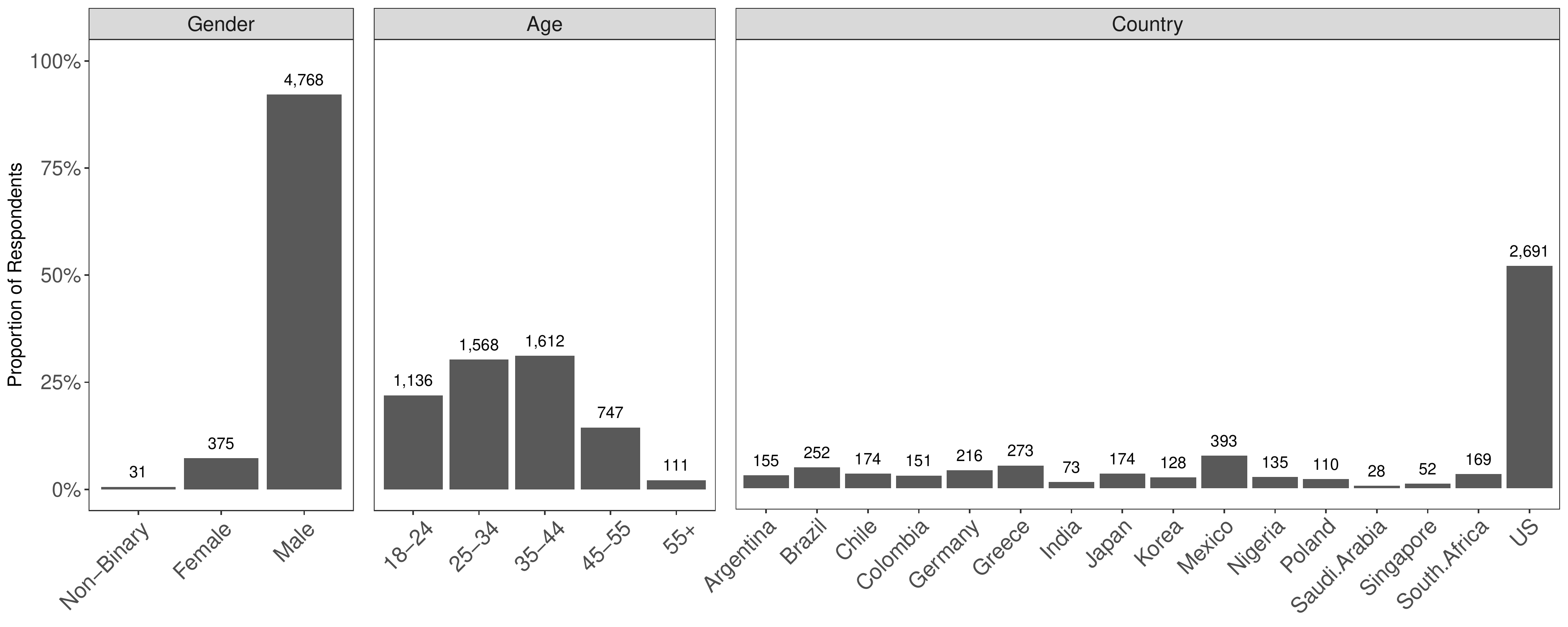}
   \caption{The distribution of survey respondents across gender, age, and country demographic subgroups. Bars correspond to percentage of group distribution; corresponding respondent counts are indicated above bars.}\label{Fig:descriptive}
\end{figure*}

\begin{table*}[h!]
\caption{The number of respondents that labeled each anonymized game. Note that one respondent can label multiple games in a survey session. Overall, 19,822 respondents at least partially answered our survey; we focus on the 5,174 respondents who completed the full survey and additionally provided full demographic information.}
\begin{tabular}{c|c|c|c|c|c|c|c|c|c|c|c}
    \toprule
    Game & 1 & 2 & 3 & 4 & 5 & 6 & 7 & 8 & 9 & 10 & 11 \\
    \midrule

     Argentina & 13 & 39 & 24 & 34 & 59 & 94 & 22 & 61 & 34 & 11 & 10 \\
     Brazil & 12 & 72 & 48 & 54 & 72 & 145 & 47 & 103 & 47 & 44 & 28 \\ 
     Chile & 27 & 37 & 19 & 41 & 52 & 99 & 61 & 72 & 33& 19 & 18 \\
     Colombia & 7 & 57 & 26 & 23 & 57 & 94 & 31 & 58 & 32 & 5 & 11 \\
     Germany & 41 & 60 & 15 & 64 & 44 & 121 & 69 & 88 & 35 & 36 & 16 \\
     Greece & 17 & 64 & 32 & 70 & 112 & 133 & 37 & 62 & 44 & 20 & 13 \\
     India & 1 & 28 & 23 & 13 & 16 & 37 & 1 & 26 & 51 & 2 & 2 \\
     Japan & 60 & 13 & 13 & 47 & 13 & 50 & 44 & 62 & 14 & 25 & 4 \\
     South Korea & 45 & 24 & 14 & 48 & 36 & 52 & 34 & 33 & 48 & 12 & 6 \\
     Mexico & 44 & 114 & 86 & 89 & 116 & 194 & 132 & 168 & 73 & 36 & 20 \\
     Nigeria & 4 & 62 & 84 & 12 & 82 & 29 & 11 & 16 & 40 & 3 & 5 \\
     Poland & 5 & 27 & 8 & 29 & 40 & 68 & 10 & 44 & 20 & 21 & 11 \\
     Saudi Arabia & 1 & 12 & 4 & 5 & 15 & 12 & 2 & 11 & 6 & 1 & 1 \\
     Singapore & 8 & 14 & 4 & 18 & 17 & 11 & 13 & 11 & 7 & 5 & 2 \\
     South Africa & 7 & 41 & 31 & 32 & 62 & 50 & 14 & 60 & 32 & 20 & 9 \\
     US & 705 & 834 & 264 & 1,125 & 203 & 1,413 & 949 & 1,376 & 576 & 695 & 238 \\
     \midrule
     \textbf{Total} & \textbf{997} & \textbf{1,498} & \textbf{695} & \textbf{1,704} & \textbf{996} & \textbf{2,602} & \textbf{1,477} & \textbf{2,251} & \textbf{1,092} & \textbf{955} & \textbf{394}\\
    \bottomrule
\end{tabular}
\end{table*}

\clearpage

\subsection{Survey Questions}\label{appendix:survey_questions}

\begin{enumerate}
    \item Are you 18 or older?
    \begin{itemize}
        Yes/No
    \end{itemize}

    \item In a typical week, how often do you play video games on each 
of the following types of devices?
\begin{itemize}
    \item Frequency Options:
    \begin{itemize}
        \item Never 
        \item Less than once per month
        \item Monthly
        \item Weekly
        \item A few times per week
        \item Daily 
    \end{itemize}
    \item Device Options:
    \begin{itemize}
        \item Personal computer
        \item Game console
        \item Tablet
        \item Smartphone
    \end{itemize}
\end{itemize}

    \item When you play video games, how often do you play them in 
English?
\begin{itemize}
   \item Frequency Options:
   \begin{itemize}
        \item Never 
        \item Less than once per month
        \item Monthly
        \item Weekly
        \item A few times per week
        \item Daily 
    \end{itemize}
\end{itemize}

    \item Please select games you played in the past year from the 
following list. We will ask questions about your selected 
games, focusing on things like gameplay and visual design. 
You are not required to select all games that you have played.
\begin{itemize}
    \item List of anonymized games
    \item None of the above
\end{itemize}

\item How complex are the controls in the game?
\begin{itemize}
\item Low complexity
\item Medium complexity
\item High complexity
\end{itemize}

\item Think back to the first time that you played the game. How many sessions did it take you to understand how to play the game?
\begin{itemize}
\item Low, 1 game session
\item Medium, 2 game sessions
\item High, 3+ game sessions, or long instruction sessions
\end{itemize}

\item How much skill does the game require?
\begin{itemize}
\item Low, new gamers would find this easy.
\item Medium, this game is medium difficulty.
\item High, this game needs significant effort and skills.
\item Extreme, punishingly difficult
\end{itemize}

\item How replayable is the game?
\begin{itemize}
\item Low, only a single play through is necessary to get most of the full experience.
\item Medium, multiple play throughs are necessary to get most of the full experience.
\item High, you can play this game an infinite number of times.
\end{itemize}

\item Now we are going to ask you about your impression of the games. It is okay if no game below satisfies the description, and it is okay if multiple descriptions are true for each game. Please check any of the boxes for the corresponding descriptions that you believe are true about each game.
\begin{itemize}
\item pacifist
\item made for kids
\item cozy
\item zen
\item a fantasy game
\item located in space
\item None of these apply
\end{itemize}

\item Please check any of the boxes for the corresponding descriptions that you believe are true about each game.
\begin{itemize}
\item centralized around the character being a hero or savior of the world
\item designed to simulate real-world activities
\item violent
\item an action game
\item None of these apply
\end{itemize}

\item Please check any of the boxes for the corresponding feelings that you had about each game.
\begin{itemize}
\item I feel strong empathy for the story or character experience.
\item I feel amused or laugh.
\item I feel that I've never played a game like this before.
\item I feel that I must strategize and make decisions to win.
\item I feel that I need to do grinding or repetitive work.
\item None of these apply.
\end{itemize}

\item Please check any of the boxes for the corresponding art styles that you believe are true about each game.
\begin{itemize}
\item an anime art style
\item an art style that invokes the memory of drawings on paper
\item a unique art style that influences the feeling of the game
\item None of these apply
\end{itemize}

\item Please check any of the boxes for the corresponding descriptions that you believe are true about each game. It is okay if no game satisfies the description, and it is okay if multiple descriptions are true for each game. Where does your fun come from when you play the game?
\begin{itemize}
\item Action: the fast-paced action, surprises, thrills, and adrenaline
\item Social: socializing and interacting
\item Mastery: mastering the skills and techniques in the game
\item Achievement: completing all the missions, increasing all the stats, or getting a more powerful character or equipment
\item Immersion: the desire to get lost in the world of the game
\item Creativity: designing your own character, house, clothes, or the world
\item None of these apply.
\end{itemize}

\item How old are you? (Optional)

\item Which gender do you identify with? (Optional)
\begin{itemize}
\item Male
\item Female
\item Another gender (please specify)
\item Prefer not to answer
\end{itemize}

\item Do you have a disability that affects how you play video games? (Optional)
\begin{itemize}
\item Yes
\item No
\item Prefer not to answer
\end{itemize}
\end{enumerate}

\clearpage

\section{Survey Responses}\label{app:survey_figs}

\begin{figure*}[!htp]
    \centering
    \includegraphics[width=\textwidth]{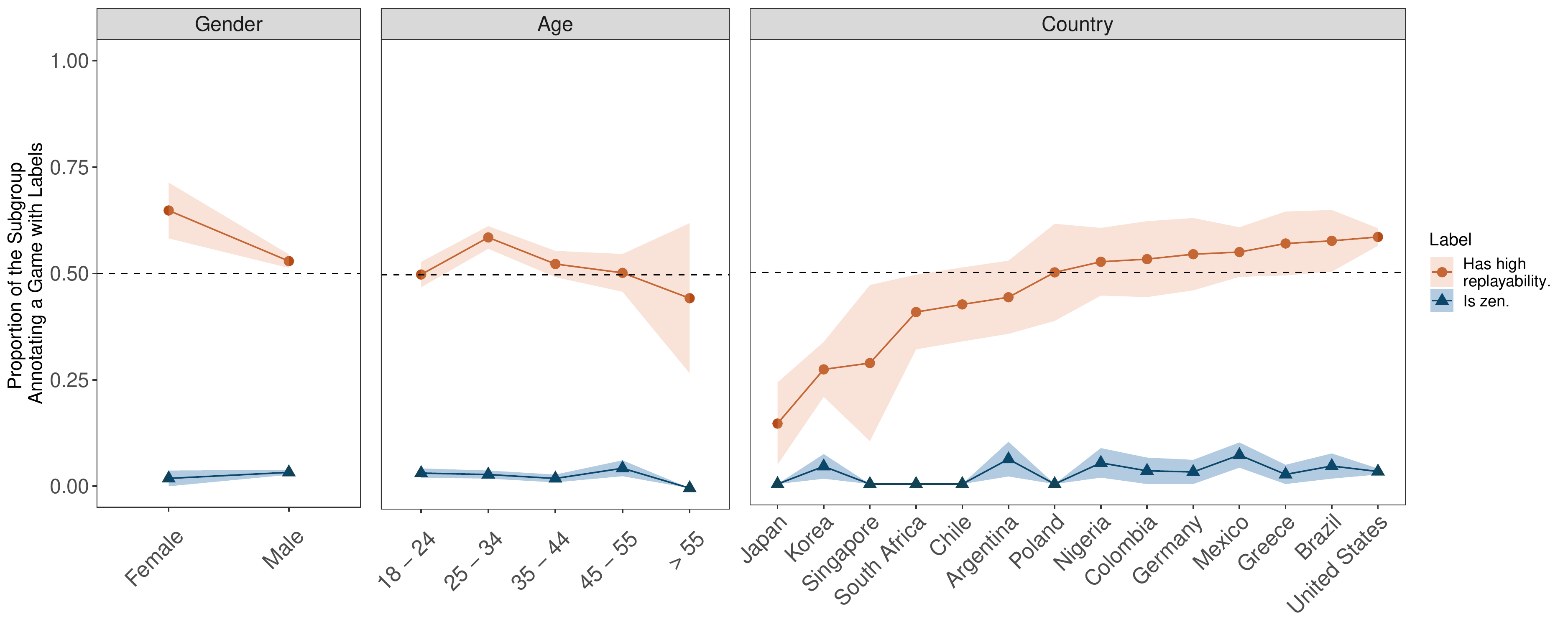}
    \caption{The proportion of the subgroup annotating a game with labels by gender, age, and countries. In contrast to the poststratified estimates in \autoref{fig:fig_2}, this plot simply shows the raw survey response data on annotation shares (with CIs presented across respondents within a specific subgroup) for the ``high replayability'' label (blue line) and ``zen'' label (orange line) for a certain action game. The similarity of this plot to that of \autoref{fig:fig_2} exemplifies that the survey respondent demographic distribution mirrors that of the underlying gamer population.}
    \label{fig:appendix_prop}
\end{figure*}

\section{Detailed Methods}\label{sec:appendixmethods}

For multi-level regression and poststratification, we used the R package \texttt{stan\_glmer} to run a logistic regression on a respondent's label annotation for a specific game, conditioned on gender and age group, and with country random effects. The \texttt{rstanarm} package uses Hamiltonian Monte Carlo; we set a 0.99 target average proposal acceptance probability for adaptation. We set the prior to be drawn from a normal distribution and the prior covariance scaled to a 0.5 exponential.

For matching analyses, we used the R package \texttt{MatchIt} with Mahalanobis distance and caliper set to 0.2. We additionally use \texttt{lmtest} and \texttt{sandwich} to calculate treatment coefficient confidence intervals and evaluate significance.

\begin{table*}[ht]
\caption{Statistics on treatment effects for language-based analysis among bilingual respondents reported in \autoref{Fig:fig_4}, for the ``makes me feel amused or laugh'' inconsistent label for a specific action game. Confidence Interval bounds are reported for 95\% CIs. Note that the $e^\beta$ column is interpreted as the a respondent being $e^\beta$ times more likely to annotate the label positively for this game when surveyed in English relative to their local language.}\label{tab:bilingual}
\begin{tabular}{c|c|c|c|c|c|c|c|c|c}
    \toprule
    Country & English & Non-English & Mean & MeanDiff & MeanDiff  &
    $e^\beta$ & $e^\beta$ & $e^\beta$ & p-value\\
     &  Mean & Mean & Difference & Lower CI & Upper CI &
     & Lower CI & Upper CI &\\
    \midrule
     Brazil & 0.37 & 0.58 & -0.21 & -0.54 & 0.12 & 0.48 & 0.11 & 2.07 & 0.33\\
     Germany & 0.81 & 0.44 & 0.38 & 0.04 & 0.71 & 6.42 & 1.39 & 29.64 & 0.02**\\ 
     Mexico & 0.44 & 0.22 & 0.22 & -0.26 & 0.71 & 1.80 & 0.11 & 28.40 & 0.68\\
     Poland & 0.25 & 0.88 & -0.63 & -1.0 & -0.18 & 0.05 & 0.00 & 0.58 & 0.02**\\ 
    \bottomrule
\end{tabular}
\end{table*}

\begin{figure*}[ht]
    \centering
    \includegraphics[width=0.9\textwidth]{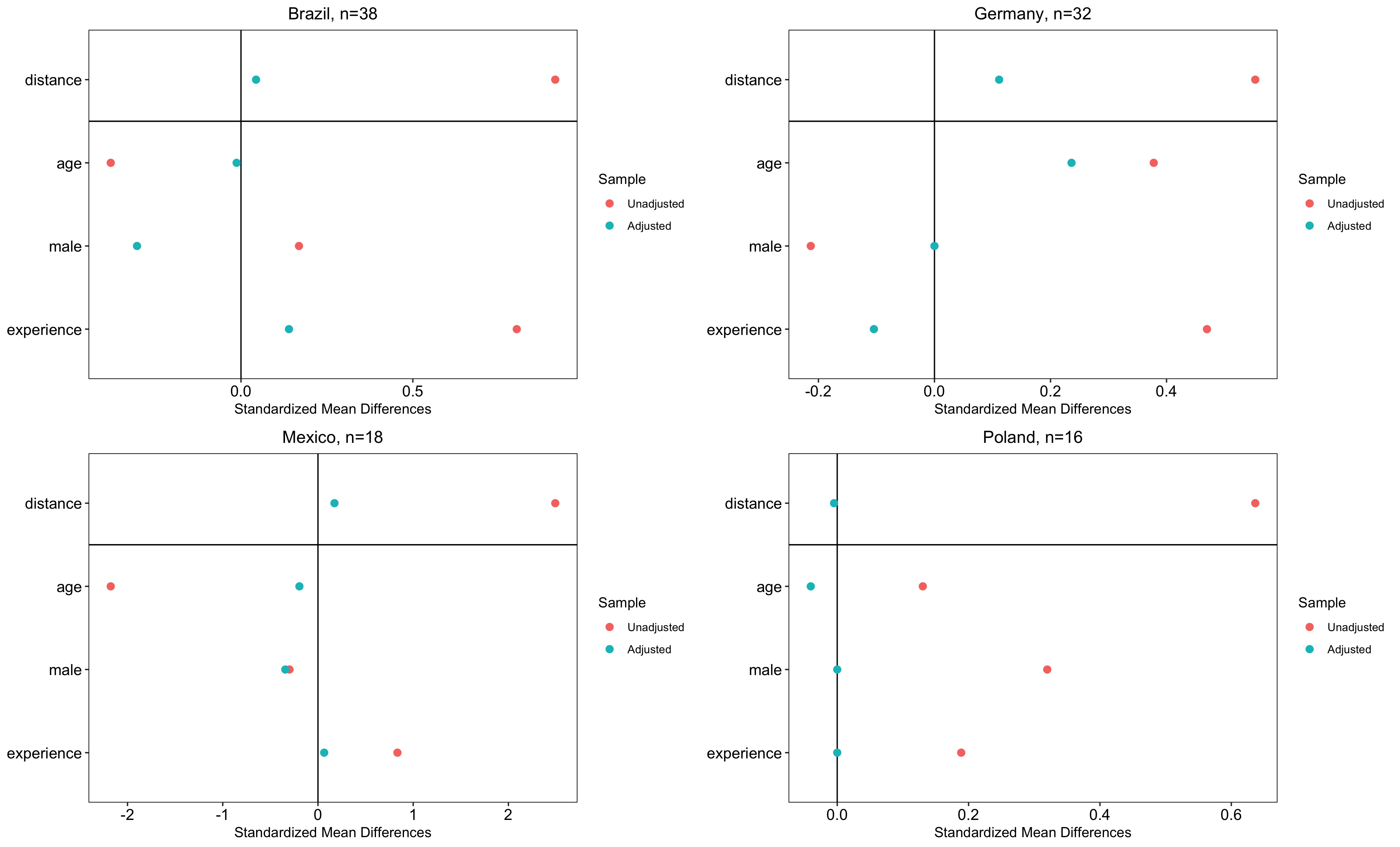}
    \caption{For a specific action game (described in \autoref{Fig:fig_4}), matching the survey response language by respondent demographic yields enough respondents in each of Brazil (n=34), Germany (n=26), Mexico (n=16), and Poland (n=16) to avoid privacy concerns. The matched subsets in each country are balanced on relevant demographic attributes.}
    \label{fig:balplots}
\end{figure*}

\begin{figure*}
    \centering
    \includegraphics[width=0.7\textwidth]{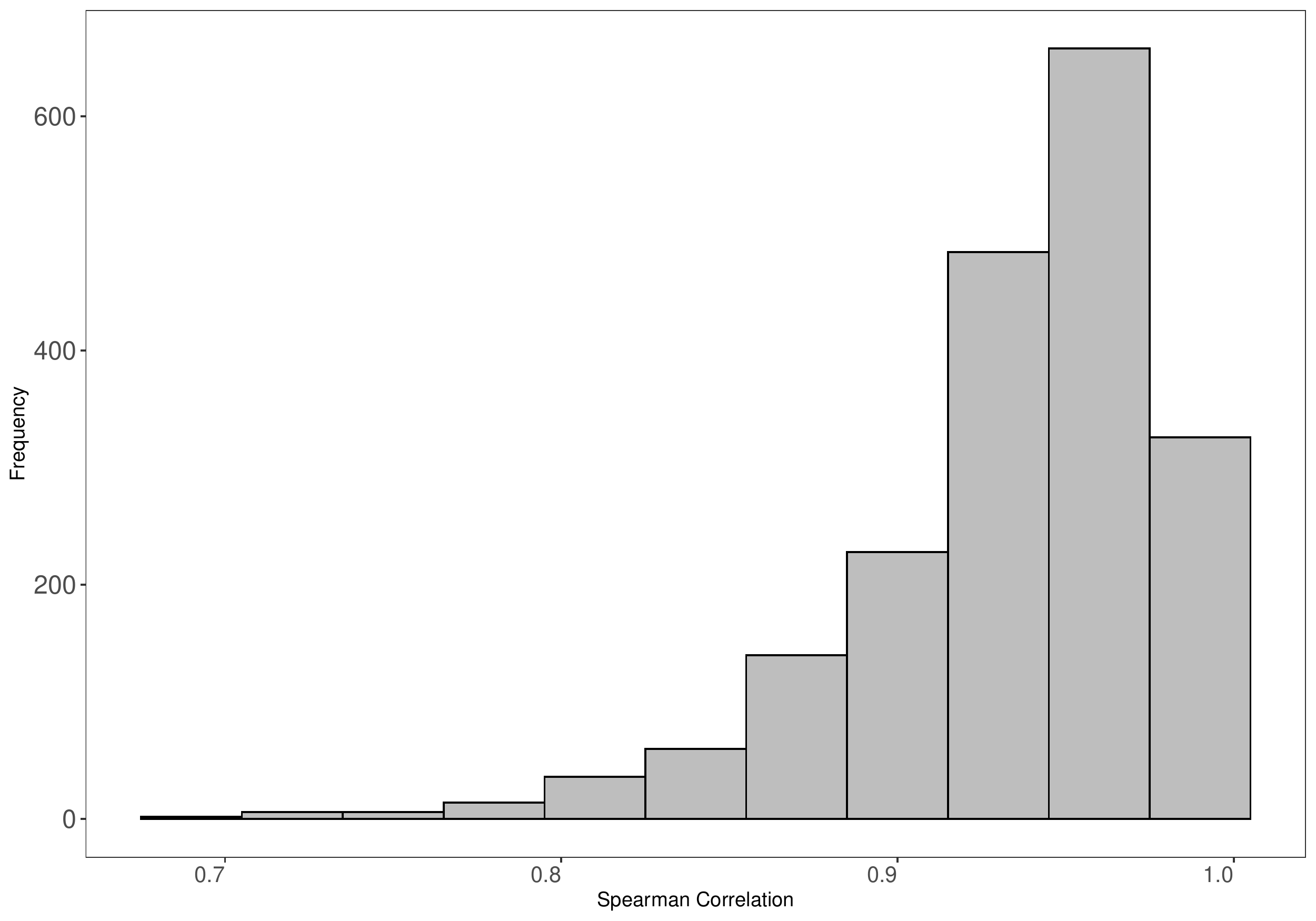}
    \caption{The distribution of the Spearman correlations (between country pairs) of ordinal label rankings in a specific game. While inconsistently conceptualized labels have significant differences across countries, the overall ordering of label ranks for a game are fairly persistent; the lowest Spearman correlation between countries is still quite high, at 0.7.}
    \label{fig:appendix_ranking_spearman}
\end{figure*}

\begin{table*}[!htbp] \centering 
  \caption{Regression results for game label annotation conditioned on Hofstede dimensions, leading to our focus on \emph{uncertainty} and \emph{orientation} as relevant cultural dimensions to study.}
  \label{tab:hofstedereg} 
\begin{tabular}{@{\extracolsep{5pt}}p{12cm}c} 
\\[-1.8ex]\hline 
\hline \\[-1.8ex] 
 & \multicolumn{1}{c}{\textit{Dependent variable:}} \\ 
\cline{2-2} 
\\[-1.8ex] & Value \\ 
\hline \\[-1.8ex] 
 GenderMale & $-$0.007 (0.014) \\ 
  AgeGroup18 - 24 & 0.132$^{***}$ (0.032) \\ 
  AgeGroup25 - 34 & 0.112$^{***}$ (0.032) \\ 
  AgeGroup35 - 44 & 0.047 (0.032) \\ 
  AgeGroup45 - 55 & $-$0.015 (0.033) \\ 
  Game2 & $-$0.407$^{***}$ (0.018) \\ 
  Game3 & $-$0.995$^{***}$ (0.024) \\ 
  Game4 & 0.354$^{***}$ (0.017) \\ 
  Game5 & $-$0.635$^{***}$ (0.020) \\ 
  Game6 & $-$0.168$^{***}$ (0.016) \\ 
  Game7 & $-$0.521$^{***}$ (0.018) \\ 
  Game7 & 0.034$^{*}$ (0.016) \\ 
  Game8 & $-$0.395$^{***}$ (0.019) \\ 
  Game9 & $-$0.143$^{***}$ (0.019) \\ 
  Game10 & $-$0.253$^{***}$ (0.026) \\ 
  Has an anime art style. & $-$2.524$^{***}$ (0.034) \\ 
  Has an art style that invokes the memory of drawings on paper. & $-$2.333$^{***}$ (0.032) \\ 
  Has high control complexity. & $-$2.196$^{***}$ (0.031) \\ 
  Has high difficulty. & $-$1.219$^{***}$ (0.026) \\ 
  Has a high learning curve. & $-$1.674$^{***}$ (0.028) \\ 
  Has high replayability. & 0.065$^{**}$ (0.024) \\ 
  Is a fantasy game. & $-$0.931$^{***}$ (0.025) \\ 
  Is an action game. & $-$0.313$^{***}$ (0.024) \\ 
  Is centralized around the character being a hero or savior of the world. & $-$1.387$^{***}$ (0.026) \\ 
  Is cozy. & $-$1.064$^{***}$ (0.025) \\ 
  Is designed to simulate real-world activites. & $-$0.703$^{***}$ (0.024) \\ 
  Is great for  designing your own character, house, clothes, or the world. & $-$0.494$^{***}$ (0.024) \\ 
  Is great for completing all the missions, increasing all the stats, or getting a more powerful character or equipment. & $-$0.146$^{***}$ (0.024) \\ 
  Is great for mastering the skills and techniques in the game. & $-$0.234$^{***}$ (0.024) \\ 
  Is great for socializing and interacting with other players. & $-$0.303$^{***}$ (0.024) \\ 
  Is great for the desire to get lost in the world of the game. & $-$0.298$^{***}$ (0.024) \\ 
  Is great for the fast-paced action, surprises, thrills, and adrenaline. & 0.036 (0.024) \\ 
  Is located in space. & $-$3.483$^{***}$ (0.048) \\ 
  Is made for kids. & $-$1.070$^{***}$ (0.025) \\ 
  Is pacifist. & $-$1.751$^{***}$ (0.028) \\ 
  Is violent. & $-$0.645$^{***}$ (0.024) \\ 
  Is zen. & $-$1.314$^{***}$ (0.026) \\ 
  Makes me feel amused or laugh. & $-$0.390$^{***}$ (0.024) \\ 
  Makes me feel strong empathy for the story or character experience. & $-$1.532$^{***}$ (0.027) \\ 
  Makes me feel that I must strategize and make decisions to win. & $-$0.364$^{***}$ (0.024) \\ 
  Makes me feel that I need to do grinding or repetitive work. & $-$0.377$^{***}$ (0.024) \\ 
  Makes me feel that I've never played a game like this before. & $-$1.747$^{***}$ (0.028) \\ 
  Power.distance & $-$0.045 (0.060) \\ 
  Individualism & $-$0.110$^{*}$ (0.045) \\ 
  Masculinity & 0.043 (0.047) \\ 
  Uncertainty & $-$0.289$^{***}$ (0.036) \\ 
  Long.term.orientation & $-$0.241$^{***}$ (0.038) \\ 
  Indulgence & 0.078 (0.044) \\ 
  Constant & 0.518$^{***}$ (0.090) \\ 
 \hline \\[-1.8ex] 
Observations & 407,512 \\ 
Log Likelihood & $-$228,518.300 \\ 
Akaike Inf. Crit. & 457,134.600 \\ 
\hline 
\hline \\[-1.8ex] 
\textit{Note:}  & \multicolumn{1}{r}{$^{*}$p$<$0.05; $^{**}$p$<$0.01; $^{***}$p$<$0.001} \\ 
\end{tabular} 
\end{table*} 

\clearpage

\section{A heatmap of the standard deviations from post-stratified estimates by country.}
\label{appendix:heatmap_country}
\begin{figure}[H]
    \centering
    \includegraphics[width=\textwidth]{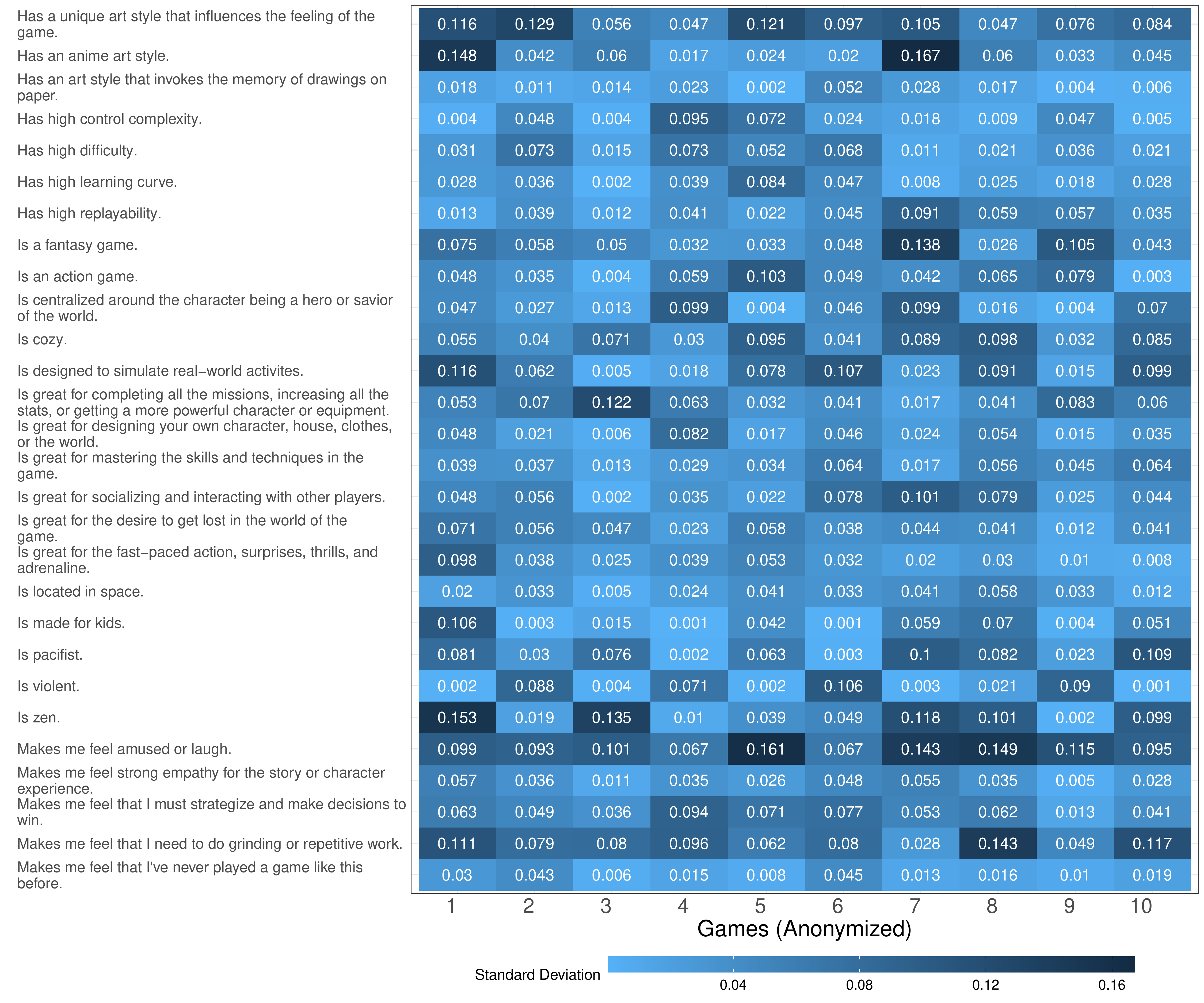}

    \caption{A heatmap of the standard deviations across countries of post-stratified label annotation estimates from \autoref{fig:fig_2}. Darker cells represent higher standard deviations and hence are more likely to be labels with inconsistent conceptualizations for a game. The surveyed games (anonymized) represent a range of popular video and mobile games across traditional video game genres.}
    \label{fig:fig_tile}
\end{figure}

\clearpage

\section{A heatmap of the standard deviations from post-stratified estimates by gender}

\begin{figure}[H]
    \centering
    \includegraphics[width=\textwidth]{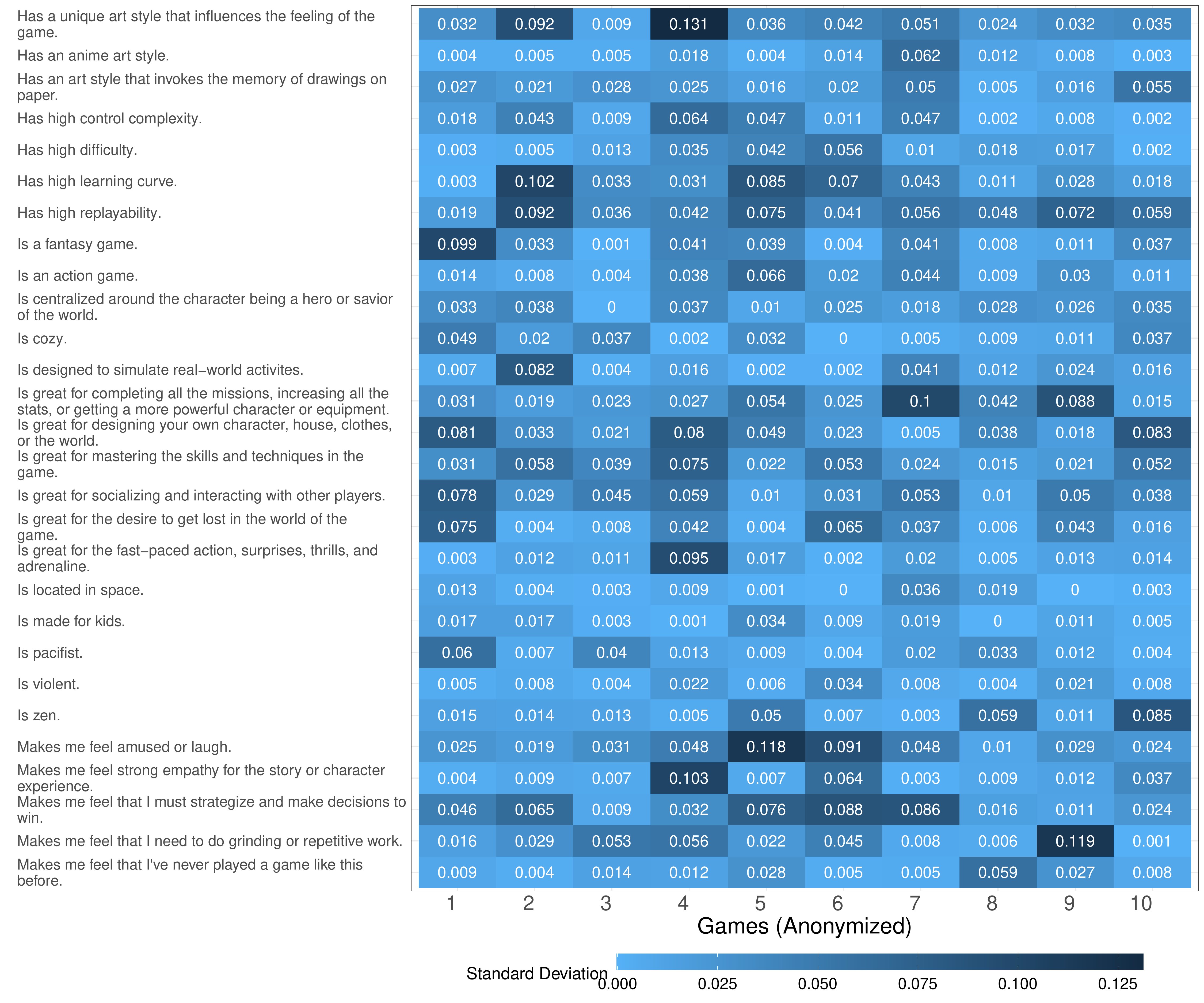}
    \caption{A heatmap of the standard deviations from MRP-adjusted estimates for gender. Darker colors represent high standard deviation. Standard deviations tend to be higher across country than across gender. The games are anonymized.}
    \label{fig:fig_gender_tile}
\end{figure}

\clearpage
\section{A heatmap of the standard deviations from post-stratified estimates by age.}

\begin{figure}[!htbp]
    \centering
    \includegraphics[width=\textwidth]{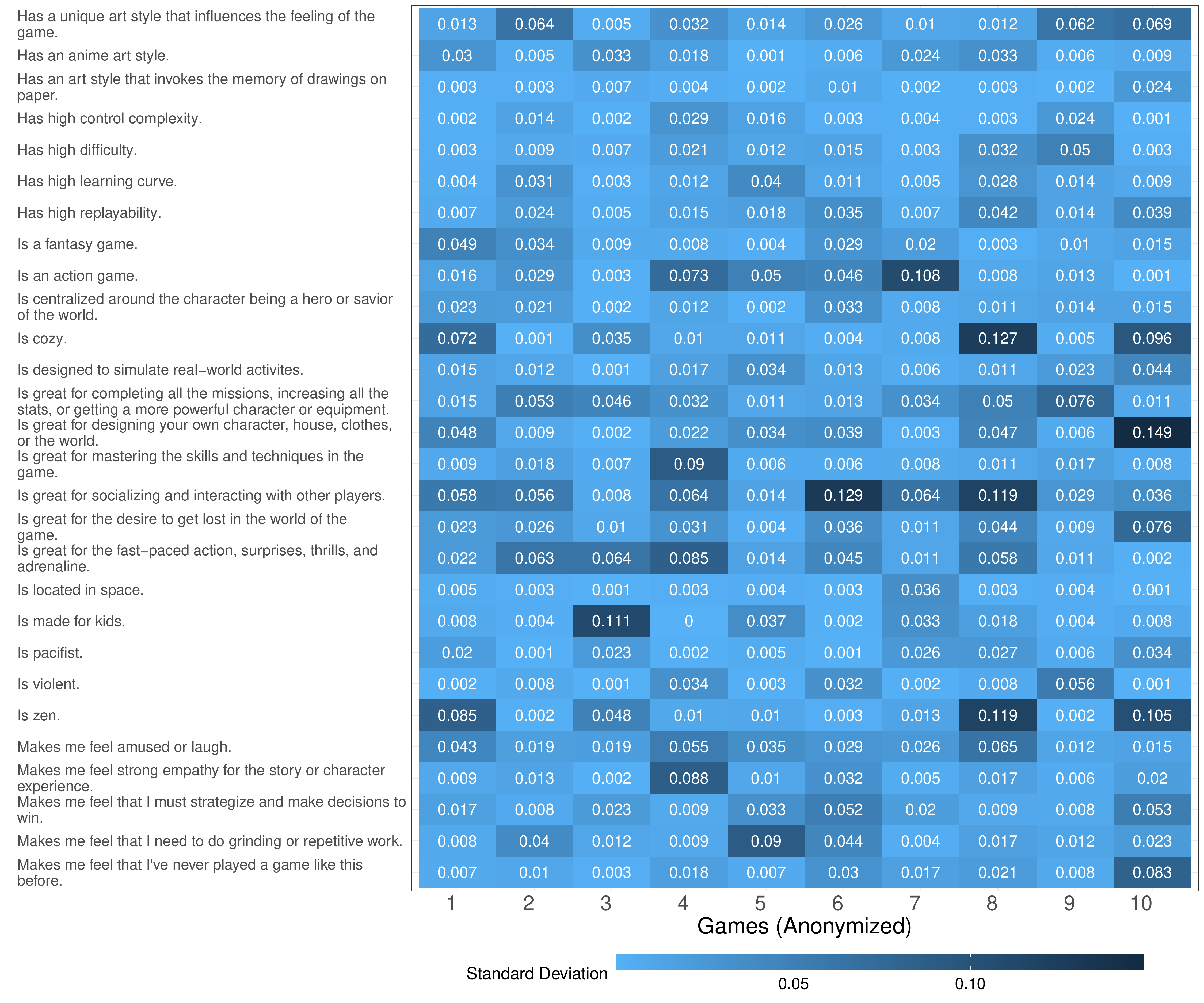}
    
    \caption{A heatmap of the standard deviations from MRP-adjusted estimates for age. Darker colors represent high standard deviation. Standard deviations tend to be higher across country than across age. The games are anonymized.}
    \label{fig:fig_age_tile}
\end{figure}

\clearpage

\section{An evaluation of F1 scores using a representative set of train and test sets.}

\begin{figure*}[!htbp]
    \centering
    \includegraphics[width=\textwidth]{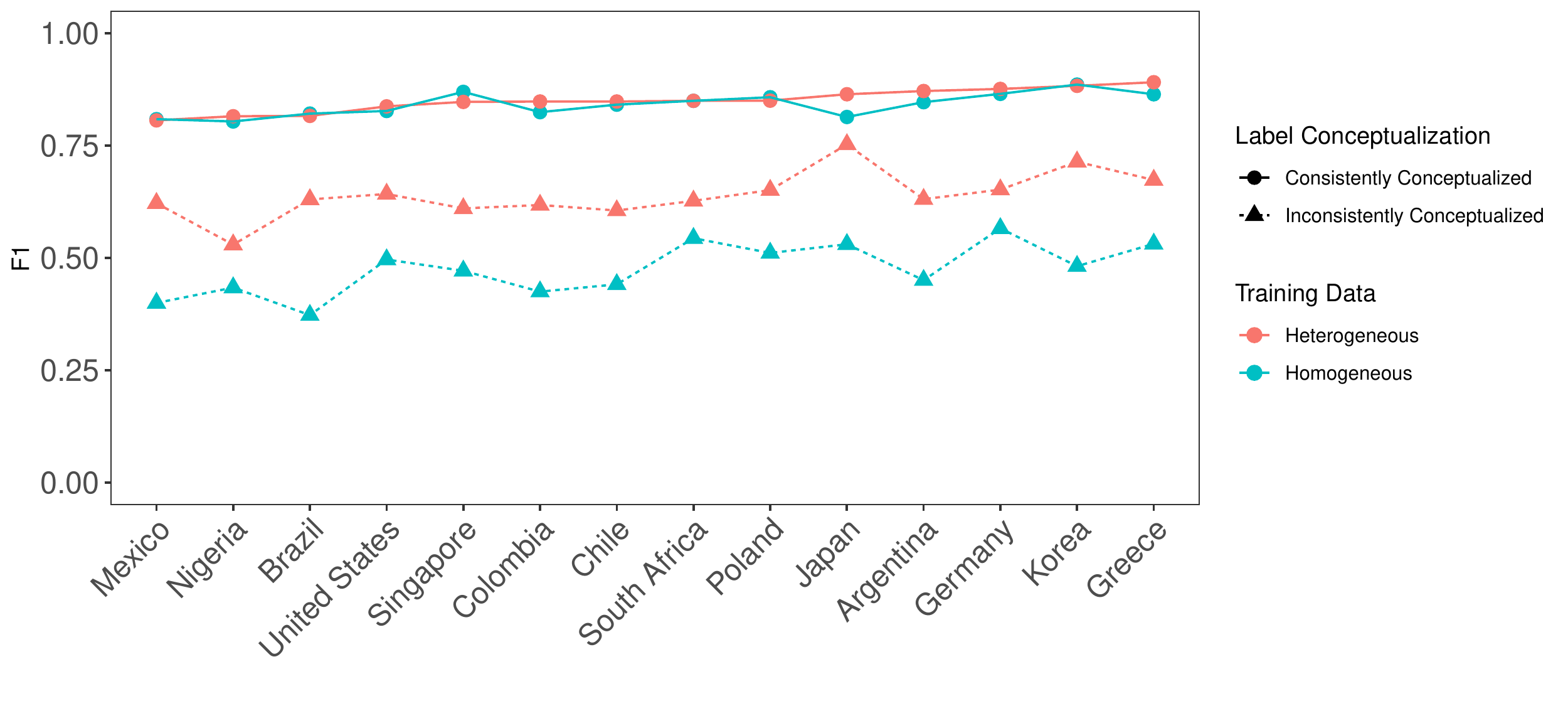}
    \caption{A robustness evaluation of F1 scores of test sets in 14 countries for logistic regression models trained on both heterogeneous and homogeneous datasets. 
    In contrast to \autoref{fig:fig_5} where train and test sets were generated using a random 70\%-30\% split of the survey data, in this figure we instead examine when the underlying homogeneous and heterogeneous datasets is representatively sampled by strata according to the overall demographic distribution of Xbox gamers (used for poststratification). Again, we find that label prediction for “inconsistently conceptualized” labels is significantly improved when models are trained on heterogeneous annotators rather than homogeneous annotators.}
    \label{fig:appendix_figure7_representative}
\end{figure*}

\begin{table*}[h]
  \caption{Comparison of evaluation metrics on models trained on homogeneous annotators versus heterogeneous annotators for binary label prediction.}
  \label{table:eval_inconsistent}
  \begin{tabular}{c c c c c c c}
    \toprule
    Label Conceptualization &
    Annotator Train Set & $\text{F1}_{\text{Train}}$ & $\text{F1}_{\text{Test}}$ & $\text{Accuracy}_{\text{Train}}$ & $\text{Accuracy}_{\text{Test}}$  \\
    \midrule
    Consistent & Homogeneous & 0.8476  & 0.8475  & 0.7608 & \textbf{0.7578} \\
    Consistent & Heterogeneous & \textbf{0.8569} & \textbf{0.8516} & \textbf{0.7642}  & 0.7585 \\
    \midrule
    Inconsistent & Homogeneous & 0.5508  & 0.5279  & \textbf{0.6597} & \textbf{0.6174} \\
    Inconsistent & Heterogeneous & \textbf{0.6263} & \textbf{0.6049} & 0.5939  & 0.6021 \\
    \bottomrule
  \end{tabular}
  \smallskip
\end{table*}